\documentstyle[12pt,amssymb,axodraw,epsfig]{article}
\renewcommand{\textfraction}{0}
\renewcommand{\topfraction}{1}

\begin{document}
\def\msbar{${\rm{\overline{MS}}}$}
\def\yhat{{\hat{y}}}
\def\oalps{${\cal{O}}(\alpha_s)$}
\setlength{\parskip}{0.45cm}
\setlength{\baselineskip}{0.75cm}
%
%
\begin{titlepage}
\begin{flushright}
DO-TH 99/02  \\
DTP/99/16    \\
February 1999 \\
\end{flushright}
\begin{center}
\vspace*{0.8cm}
{\LARGE
\hbox to\textwidth{\hss
{\bf Next-to-Leading Order QCD Corrections to} \hss}

\vspace{+0.2cm}
\hbox to\textwidth{\hss
{\bf Charged Current Charm Production and the \hss}}

\vspace{+0.2cm}
\hbox to\textwidth{\hss
{\bf Unpolarized and Polarized Strange Sea at HERA} \hss}}

\vspace*{1.0cm} 
{\large S.\ Kretzer}\\
{Institut f\"{u}r Physik, Universit\"{a}t Dortmund,
D-44221 Dortmund, Germany} \\

\vspace*{0.5cm}
and\\

\vspace*{0.5cm}
{\large M.\ Stratmann}\\
{Department of Physics, University of Durham, Durham DH1 3LE,
England}\\
\vspace*{1.6cm}
{\bf Abstract}

\vspace{-0.3cm}
\end{center}
%
Charged current charm and $D$ meson production is studied in detail as a 
means of determining the unpolarized and polarized strange sea densities at HERA.
All analyses are performed in next-to-leading order QCD, including a 
calculation of the so far unknown spin-dependent \msbar\
coefficient functions up to \oalps.
It is shown that a decent measurement is possible in the unpolarized
case, provided a sufficient luminosity can be reached
in the future, while for longitudinally polarized beams it appears to 
be extremely challenging due to limitations imposed on the
expected statistical accuracy by the charm detection efficiency.
\end{titlepage} 
%
%
\section{Introduction}
At experimentally relevant $Q^2$ values the different unpolarized
flavor sea quark distributions $\bar{q}(x,Q^2)$,
$\bar{q}=\bar{u},\,\bar{d},$ and $\bar{s}$, are quite distinct,
and only for asymptotically large values of $Q^2$ they eventually
evolve to a common $x$ shape due to the dominance of 
$g\rightarrow q\bar{q}$ transitions. These light quark sea distributions
are usually treated as massless partons in all sets of parton 
densities, hence requiring some non-perturbative input 
for their $Q^2$ evolutions which has to be determined experimentally.
Heavy quarks ($m_{\bar{q}}\gg\Lambda_{QCD}$, i.e., $\bar{q}=\bar{c}$
and $\bar{b}$), however, can be dealt with purely perturbatively,
within different methods though, which completely determines their 
$x$ and $Q^2$ dependence, with the heavy quark masses $m_{\bar{q}}$ 
being the only physical parameters.

In recent sets of unpolarized parton distributions it is either 
assumed that $\bar{s}$ $(=s)$ has the same $x$ shape as $\bar{u}+\bar{d}$ 
\cite{ref:mrst,ref:cteq4}, or $\bar{s}$ is solely generated
by QCD dynamics from a vanishing input distribution at some low
scale \cite{ref:grv98}, in both cases leading to an overall
suppression of the $x$ integrated second moment $\int dx\, x\bar{s}$ 
in the light sea of about $50\%$ at $Q^2\simeq 5-10\,\mathrm{GeV}^2$ [4-7],
presumably due to the larger mass of strange quarks. 
The entire $x$ dependence of the flavor decomposed
unpolarized light sea is, however, still rather uncertain.
While some information about $\bar{u}$ and $\bar{d}$ is now available 
from various sources\footnote{Recent compilations can be found, for instance, in 
Refs.~\cite{ref:mrst,ref:grv98}.}, and all data indicate that $\bar{d}$ 
is greater than $\bar{u}$, $\bar{s}$ can be inferred only from CCFR data on 
deep inelastic neutrino charm production \cite{ref:ccfrlo,ref:ccfrnlo} for the 
time being, with a NuTeV update to be expected in the near future \cite{ref:adams}.
An alternative extraction along
$\frac{5}{6} F_2^{\nu N}(x,Q^2)- 3 F_2^{\mu N}(x,Q^2) \simeq
 x \bar{s}(x,Q^2)$ (or equivalently from corrections to the 
$F_2^{\mu N}/ F_2^{\nu N}\simeq 5/ 18$ rule)
combining CCFR \cite{ref:ccfrsf} and NMC \cite{ref:nmcsf} data cannot 
be reliably performed because it suffers from the fact that $\bar{s}$ 
emerges only as a small residual of two large numbers 
($x \bar{s}\ll F_2^{\mu N}, F_2^{\nu N}$), which 
furthermore appear to be incompatible at low $x$ \cite{ref:csb}. 
At present, results for  $\frac{5}{6} F_2^{\nu N} - 3 F_2^{\mu N}$
seem to be in conflict with the CCFR charm production data
\cite{ref:csb,ref:gkr1}, and if this tendency persists with future
NuTeV data, it requires further clarification \cite{ref:gkr1}.

The leading order (LO) contribution to charged current (CC)
charm production in deep inelastic scattering (DIS) is given by the
${\cal{O}}(\alpha_s^0)$ parton model process
\begin{equation}
\label{eq:loproc}
W^+ s' \rightarrow c\;\;,
\end{equation}
depicted in Fig.~1(a), where $s'$ denotes the Cabibbo-Kobayashi-Maskawa (CKM) 
`rotated' combination
\begin{equation}
\label{eq:ckm}
s' \equiv \left|V_{cs}\right|^2 s +
          \left|V_{cd}\right|^2 d
\end{equation}
with $|V_{cs}|=0.9745$ and $|V_{cd}|=0.2205$ \cite{ref:pdg}. Due
to the smallness of $\left|V_{cd}\right|^2$ in (\ref{eq:ckm}) the
process (\ref{eq:loproc}) is expected to be essentially sensitive to the 
strange sea content.
Only at large $x$, where quark sea contributions are less relevant, 
the $\left|V_{cd}\right|^2$ suppression is balanced by the 
valence enhancement of the well-known $d_v(x)$. 

In next-to-leading order (NLO) QCD, however, this simple picture is
spoiled, and the complete set of Feynman diagrams in Fig.~1 has
to be considered.
Apart from the virtual and real \oalps\ corrections to (\ref{eq:loproc}), 
the genuine NLO gluon induced subprocess
\begin{equation}
\label{eq:nloproc}
W^+ g \rightarrow c \bar{s}'
\end{equation}
has to be taken into account as well, which may yield a significant
contribution \cite{ref:aot} to the charm production cross section, hence 
representing an important `background' for any extraction of the strange sea.
In case of inclusive charm production these NLO corrections
have been calculated for unpolarized target nucleons
in different regularization prescriptions such as the conventional \msbar\ scheme 
\cite{ref:gottschalk,ref:gkr1}, which we henceforth adopt, or the ACOT scheme 
\cite{ref:acot,ref:kschie1}, which also takes into account possible effects 
of a finite strange quark mass. Recently these fully inclusive
calculations were extended to the experimentally relevant case of momentum 
$(z)$ distributions of the produced $D$ mesons [18-20]. 
Such detailed production cross section considerations seem to prefer
a softer, `heavy quark-like', strange sea \cite{ref:grv98} over 
${\bar s} \propto {\bar u} + {\bar d}$ inputs \cite{ref:ccfrnlo,ref:mrst,ref:cteq4},
but further experimental clarification is certainly highly desirable.

Besides new fixed target neutrino data from NuTeV \cite{ref:adams}, 
one interesting possibility to shed more light on the strange
density and its $Q^2$ evolution would be, of course, to study CC charm,
i.e., dominantly $D$ meson, production in $e^{\pm}p$ collisions at HERA, 
provided a sufficient luminosity can be reached. We shall perform a closer 
analysis of this option and the impact of the gluonic `background' 
(\ref{eq:nloproc}) in this case below.  
It should be noted in passing that a high precision measurement
of CC charm production in $e^-p$ {\em and} $e^+p$ collisions could
possibly reveal also the relevance of recent claims \cite{ref:brodsky}
that strange and anti-strange densities may differ, i.e., $s\neq \bar{s}$, 
contrary to what is assumed in all analyses of parton densities so far.

Turning to longitudinally polarized parton densities $\Delta f$,
defined by
\begin{equation}
\label{eq:poldef}
\Delta f(x,Q^2) \equiv f_+(x,Q^2) - f_-(x,Q^2)\;\;,
\end{equation}
where $f_+$ $(f_-)$ denotes the distribution of a parton $f$ with 
its spin (anti-)aligned to the parent nucleon's 
spin\footnote{By taking the sum instead of the difference in 
(\ref{eq:poldef}) one recovers the unpolarized (helicity-averaged)
parton densities $f$.}, much less is known experimentally about the flavor 
decomposition of the light sea or even the gluon density $\Delta g$. 
Information on the $\Delta f$ is so far almost exclusively available from
fully inclusive polarized DIS, i.e., structure function measurements 
\cite{ref:polover}, which are only sensitive to specific non-singlet 
and singlet combinations of the spin-dependent quark densities and not to a 
full flavor 
separation or $\Delta g$.
Thus all current sets of polarized parton densities, 
such as, e.g., the GRSV \cite{ref:grsv} or GS \cite{ref:gs} distributions, 
have to fully rely on certain assumptions when providing flavor decomposed quark 
densities, which are often biased by unpolarized measurements and, of course, 
remain to be checked.

The knowledge of the flavor decomposed polarized densities, more
specifically of their first moments $\Delta f(Q^2)$
(as obtained by taking the $x$ integral in (\ref{eq:poldef})), 
is moreover required to understand how the nucleon's spin $S_z^N=1/2$ is shared 
by its `constituents' as a function of the momentum transfer $Q^2$,

\begin{equation}
\label{eq:spinsum}
S_z^N = \frac{1}{2} = \frac{1}{2} \sum_{q=u,d,s} \left[ 
\Delta q(Q^2)+\Delta \bar{q}(Q^2)\right]
 + \Delta g(Q^2) + L_z^q(Q^2) + L_z^g(Q^2)\;\;,
\end{equation}
where $L_z^q$ $(L_z^g)$ denotes the orbital angular momentum contribution
of the quarks (gluons)\footnote{Of course, in NLO the decomposition
on the right-hand side (r.h.s.) of (\ref{eq:spinsum}) 
becomes factorization scheme dependent, and one always has to 
specify the scheme one is referring to when quoting values for the first moments
$\Delta f(Q^2)$ or $L_z^{q,g}$.}.

The presently available semi-inclusive spin-dependent DIS measurements 
\cite{ref:semiincl} are still not conclusive enough to disentangle different flavors
reliably, but some progress has to be expected soon in particular from
the HERMES experiment.
Together with upcoming measurements of $W$ boson production 
at the polarized BNL-RHIC $pp$ collider this may yield some information 
about $\Delta \bar{u}$ and $\Delta \bar{d}$. However, a direct measurement 
of $\Delta \bar{s}$ like in the CCFR neutrino DIS experiment turns out to be 
extremely remote despite that neutrinos have definite helicity, since 
tons of target material would have to be polarized. Thus other
possibilities have to be examined here. Since it is no longer inconceivable that 
HERA can be operated at some stage in the future in a longitudinally 
polarized collider mode \cite{ref:herapol}, it was suggested to study CC 
charm production to decipher $\Delta \bar{s}$ \cite{ref:maul}
along similar lines as discussed above for unpolarized $e^{\pm}p$ collisions.
However, these studies neither have been performed in NLO, nor do they 
include a realistic estimate of the expected statistical or theoretical uncertainties 
for such a measurement at a polarized HERA. 
It is the main purpose of this paper to provide the complete NLO framework 
for CC inclusive charm and momentum $(z)$ differential $D$ meson production 
in the \msbar\ scheme\footnote{We compare our calculation to existing
inclusive \msbar\ \cite{ref:svw} as well as $k_T^{\min}$- \cite{ref:vw} and 
mass-regulated \cite{ref:schienbein} results at the end of Section 2.}  
with polarized beams and to study the prospects of a measurement of 
$\Delta \bar{s}$ at a polarized HERA.  

The remainder of the paper is organized as follows: in Section 2 we
outline all relevant technical details of the calculation of 
unpolarized and polarized CC charm production in NLO, mainly focusing on additional 
complications which arise in the momentum $(z)$ differential and in the 
spin-dependent case. Section 3 is devoted to a detailed numerical analysis. First we
discuss the prospects of a determination of the unpolarized
strange density at HERA, then we turn to the polarized HERA option
and its potential of learning something about $\Delta \bar{s}$. 
The Appendix contains the polarized coefficient functions,
which are our main analytical results and are too long to be presented
in the text.

\section{Technical Framework}
To derive the cross sections for inclusive charm and momentum $z$ differential 
$D$ meson production in longitudinally polarized CC DIS we follow closely 
the corresponding unpolarized NLO calculations in 
\cite{ref:gottschalk,ref:gkr1,ref:gkr2}. Since 
only very few technical details have been presented in \cite{ref:gkr2}
in the experimentally more relevant case of heavy meson production, we shall 
give a brief overview of the most important, non-trivial calculational steps as well.
We mainly focus, however, on the complications arising
due to the appearance of $\gamma_5$ and the Levi-Civita tensor 
$\epsilon_{\mu\nu\rho\sigma}$ in course of the calculations, 
which requires special attention in the polarized case as we will discuss below.

The NLO corrections to the LO parton model CC production mechanism\footnote{
For simplicity we ignore the CKM $s\leftrightarrow d$ mixing (\ref{eq:ckm}) 
in this technical section. It is, however, properly taken into account in all 
phenomenological applications in the next section.} (\ref{eq:loproc}) stem from 
the boson gluon fusion (BGF) and real gluon emission subprocesses (\ref{eq:nloproc}) 
and $W^+ s \rightarrow c g$, respectively. In the latter case
also virtual corrections to (\ref{eq:loproc}) have to be included. All
contributions are represented by their Feynman diagrams in Fig.~1. 
As usual, the NLO (\oalps) diagrams comprise soft and collinear
divergences, and one has to choose a consistent method of regularizing
these singularities so that they become manifest.
For this purpose we work in the well established framework of
dimensional regularization in $n=4+2\varepsilon$ spacetime dimensions.
The divergences then occur as poles $\sim 1/\varepsilon$ and 
$\sim 1/\varepsilon^2$ in the physical limit $n\rightarrow 4$.
The latter double pole terms only arise in the quark initiated 
subprocess when soft and mass/collinear singularities coincide.
While these double and the single poles from soft gluons in virtual loops and 
real soft gluon emission have to cancel by the KLN theorem, 
there remain mass/collinear poles $\sim 1/\varepsilon$ in the gluon and quark 
initiated NLO corrections stemming from the region in
phase space where a strange quark propagator goes on-shell.
This can happen either when the initial state gluon splits into a collinear 
$s {\bar{s}}$ pair or when the initial state strange quark radiates off a 
collinear real gluon. These poles have to be removed from the production dynamics by 
factorizing them off into the renormalized (scale-dependent) strange
sea density, where we adopt the commonly used \msbar\ factorization prescription. 
Charm quark propagators in collinear $g\rightarrow c {\bar{c}}$, $c\rightarrow c g$ 
subdiagrams are protected from going on-shell by the (heavy) charm quark mass, 
which thereby acts as an effective cut-off of non-perturbative long distance strong 
interactions. 

In the following we give a unified description of the calculation of the 
relevant production processes for unpolarized and longitudinally polarized initial 
states. The possibility to obtain the unpolarized results `simultaneously'
provides a useful check of the correctness of our results, and we fully agree with 
the unpolarized results presented in \cite{ref:gottschalk,ref:gkr1,ref:gkr2}.
To be specific, we calculate the contributions of incoming quarks and
gluons to the unpolarized and polarized structure functions 
$H_i^{q,g}$ and $\Delta H_i^{q,g}$, respectively, as depicted in Fig.~1
by properly projecting the helicity dependent squared matrix elements
$|{\cal{M}}_{\alpha\beta}^{q,g}(\pm)|^2$ onto the structure function $i$: 
\begin{equation}
\label{eq:polunpol}  
\left\{ H_i^{q,g} \atop \Delta H_i^{q,g} \right\}
\equiv \left\{ P^{\alpha\beta}_i \atop \Delta P^{\alpha\beta}_i \right\}
\left[ \left| {\cal{M}}_{\alpha\beta}^{q,g}(+) \right|^2 \pm 
       \left| {\cal{M}}_{\alpha\beta}^{q,g}(-) \right|^2 
\right] d{\mathrm{PS}}_2\;\;. 
\end{equation}
The indices $\alpha$ and $\beta$ in (\ref{eq:polunpol}) indicate
the polarization indices of the $W^{\pm}$ boson, and $d{\mathrm{PS}}_2$ is the 
two body phase-space as defined in (\ref{eq:dps2}) below. The operators 
$P_i^{\alpha\beta}$ are given in Eqs.~(B9)-(B11) of \cite{ref:gottschalk} and 
project for $i=1,\,2,$ and 3 onto the relevant unpolarized structure functions 
$F_1$, $F_2$, and $F_3$, respectively.
The structure functions $F_4$ and $F_5$ in \cite{ref:gottschalk} do not
contribute to the lepton-nucleon CC cross section if one assumes a
vanishing mass for the lepton as we will do in the following. Since the polarized 
structure functions appear in a similar way in the hadronic tensor as the 
unpolarized ones (cf.\ also Eqs.~(\ref{eq:unpolxsec}) and (\ref{eq:polxsec}) below), 
the same projection operators apply in this case if one identifies
\begin{equation}
\label{eq:polproj}
\Delta P_1^{\alpha\beta}\equiv P_3^{\alpha\beta}\;,\;
\Delta P_3^{\alpha\beta}\equiv P_1^{\alpha\beta}\;,\;
\Delta P_4^{\alpha\beta}\equiv P_2^{\alpha\beta}\;\;.
\end{equation}
The operators $\Delta P_i^{\alpha\beta}$ in (\ref{eq:polproj}) then project onto 
the relevant polarized structure functions $g_1$, $g_3$, and $g_4$, respectively,
and other structure functions like $g_6$ and $g_7$ again do not contribute for a 
vanishing lepton mass \cite{ref:vw}. The projection onto the helicity states 
$h=\pm$ of the incoming strange quark or gluon in the matrix elements ${\cal{M}}$ 
in (\ref{eq:polunpol}) is achieved by using the standard relations 
(see, e.g., \cite{ref:craigie})
\begin{equation}
\label{eq:quaproj}
u(p_s,h) \bar{u}(p_s,h) = \frac{1}{2} \not\!  p_s (1-h \gamma_5)
\end{equation}
for incoming strange quarks with momentum $p_s$ (analogously for anti-strange quarks) 
and
\begin{equation}
\label{eq:gluproj}
\epsilon_{\mu}(p_g,h) \epsilon^*_{\nu}(p_g,h)
= \frac{1}{2} \left[ -g_{\mu\nu} + i h \epsilon_{\mu\nu\rho\sigma}
\frac{p_g^{\rho}q^{\sigma}}{p_g\cdot q}\right]
\end{equation}
for incoming gluons with momentum $p_g$ ($q$ denotes the four-momentum of the 
$W^\pm$ boson).

The presence of $\gamma_5$ and the totally anti-symmetric tensor
$\epsilon_{\mu\nu\rho\sigma}$ in the V-A vertices and
$P_3^{\alpha\beta}=\Delta P_1^{\alpha\beta}$, respectively, and
-- in the polarized calculation -- also due to (\ref{eq:quaproj}) and 
(\ref{eq:gluproj}), introduces some extra
complications, because their purely four-dimensional origin allows for
no straightforward continuation to $n\ne 4$ dimensions. We choose to handle these 
quantities in the HVBM scheme \cite{ref:hvbm}, which was shown to provide an 
internally consistent continuation of $\gamma_5$ and $\epsilon_{\mu\nu\rho\sigma}$ 
to arbitrary dimensions. This prescription is also implemented in the 
package {\tt{TRACER}} \cite{ref:tracer}, which we use for all trace calculations 
in $n$ dimensions. In the HVBM scheme \cite{ref:hvbm} the $\epsilon$-tensor 
continues to be a genuinely four-dimensional object, and $\gamma_5$ is defined as
in four dimensions, implying $\{\gamma^{\mu},\gamma_5\}=0$ for
$\mu=0,\,1,\,2,\,3$ and $[\gamma^{\mu},\gamma_5]=0$ otherwise.
This effectively splits the $n$ dimensional space into two subspaces: one
containing the four space-time dimensions and one containing the remaining
$(n-4)$ dimensions, denoted as `hat-space' henceforth. The price to pay is that 
in the matrix elements squared in (\ref{eq:polunpol}) we then encounter not only 
conventional $n$ dimensional scalar products of two momenta, 
which can be expressed in terms of the usual partonic Mandelstam variables,
\begin{equation}
\label{eq:partmandel}
s=(p_{s,g}+q)^2\;\;,\;\;t=(q-p_c)^2\;\;,\;\;u=(p_{s,g}-p_c)^2
\end{equation}
in our case, but also similar scalar products in the hat-space. However, 
in the parton-boson c.m.\ system with the incoming momenta chosen to lie in the 
$\pm z$ direction, all possible $(n-4)$ dimensional scalar products of the two 
final state momenta can be expressed in terms of a single hat momenta 
combination $\hat{k}^2=-\widehat{k\cdot k}$ due to momentum conservation.
The $\hat{k}^2$ terms do not contribute to the unpolarized 
calculations\footnote{In principle these calculations can be performed also by 
using the theoretically inconsistent anti-commuting $\gamma_5$ prescription 
of \cite{ref:chanowitz} in $n$ dimension as was done in \cite{ref:gottschalk}.} 
\cite{ref:gottschalk,ref:gkr1,ref:gkr2}, but are important for polarized DIS due 
to the additional appearance of $\gamma_5$ and $\epsilon_{\mu\nu\rho\sigma}$ in 
Eqs.~(\ref{eq:quaproj}) and (\ref{eq:gluproj}) as we will discuss in more detail 
next.

The partonic two particle phase space $d\mathrm{PS}_2$ for one massive and
one massless parton in (\ref{eq:polunpol}) is given by
\begin{eqnarray} 
\label{eq:dps2}  
\int d{\mathrm{PS}_2} &=& \frac{1}{8\pi}\ (4\pi)^{-\varepsilon}
\frac{s-m_c^2}{s}
\ \frac{1}{\Gamma(\varepsilon)}
\ \int_0^1\ d{\hat{y}} 
\ \int_0^{\frac{(s-m_c^2)^2}{s} {\hat{y}} (1-{\hat{y}})}
\ d\hat{k}^2 
\ \left(\hat{k}^2\right)^{-(1-\varepsilon)}  
\\ 
\label{eq:dps2nohat}  
&=& \frac{1}{8\pi}\ \frac{s-m_c^2}{s}
\ \frac{1}{\Gamma(1+\varepsilon)}
\ \left[ \frac{(s-m_c^2)^2}{4\pi s} 
\right]^\varepsilon\ \int_0^1\ \left[{\hat{y}}(1-{\hat{y}})
\right]^\varepsilon\ d{\hat{y}} \ \ \ .
\end{eqnarray}   
In (\ref{eq:dps2nohat}) the integration over $d\hat{k}^2$ has been carried out,
and the standard $n$ dimensional phase space \cite{ref:gottschalk} is recovered. 
This can {\em only} be done if either the matrix element squared 
${\cal{\left|M\right|}}^2$ in (\ref{eq:polunpol}) trivially does not depend on 
$\hat{k}^2$ or if the $\hat{k}^2$ dependent terms in ${\cal{\left|M\right|}}^2$ 
are not multiplied by a sufficiently singular term $\sim (1-\hat{y})^{-2}$, because 
the extra subvolume of the phase space, which is available for $\hat{k}^2$, is 
intrinsically of order ${\cal{O}}(\varepsilon)$ due to the $1/\Gamma(\varepsilon)$ 
in (\ref{eq:dps2}). These conditions are always met in the unpolarized case.

Eq.~(\ref{eq:dps2}) leaves the two options to either fully integrate
the $(\Delta)H_i^{q,g}$ in (\ref{eq:polunpol}) over the entire phase space or to 
consider more differential observables which can be obtained from some Jacobian 
according to $d/d \Xi = d/d {\hat{y}}\ d{\hat{y}}/d\Xi$, where $\Xi$ stands for any 
kinematical observable that can be expressed by 
${\hat{y}}\equiv (1+\cos \theta^\ast )/2$, with $\theta^\ast$ being the 
$W^{\pm}$-parton c.m.s.\ scattering angle. A more detailed discussion including 
subtleties arising from endpoint (soft) singularities will be given below when we 
consider the fragmentation spectrum of charm quarks in CC DIS. 

The helicity-dependent matrix elements 
$\left| {\cal{M}}^{q,g}_{\alpha\beta}(\pm)\right|^2$ in
(\ref{eq:polunpol}) can be easily derived from standard
Feynman rules and will not be given here\footnote{The non-trivial 
virtual corrections are explicitly calculated in Ref.~\cite{ref:gottschalk}
which we confirm after eliminating a misprint in the coefficient
$A_2$ in Tab.~1 of \cite{ref:gottschalk}, which should read
\cite{ref:svw} $K_A$ instead of $K_A/2$.}.
They can be conveniently expressed in terms of the partonic Mandelstam variables
(\ref{eq:partmandel}), which in turn can be written as 
\begin{eqnarray} 
\nonumber
s &=& \frac{Q^2}{\xi^\prime}\ 
\left(1-\xi^\prime+\frac{m_c^2}{Q^2}\right)\\
\label{eq:mandelstams}
t &=& \frac{-1}{s}\ 
(s+Q^2)(s-m_c^2)(1-{\hat{y}})\\
\nonumber
u &=& -\frac{s+Q^2}{s}\left[m_c^2+{\hat{y}}
(s-m_c^2)\right]+ m_c^2\;\; ,
\end{eqnarray} 
where $Q^2=-q^2$ and
\begin{equation}
\xi^\prime = \frac{Q^2}{2p_{s,g}\cdot q}\ 
\left(1+\frac{m_c^2}{Q^2}\right)=\frac{Q^2+m_c^2}{s+Q^2}
\end{equation}
is the partonic analogue of the slow rescaling parameter $\xi=x(1+m_c^2/Q^2)$ 
\cite{ref:barnett}. Its introduction is required in 
NLO for a consistent factorization of initial state collinear singularities 
\cite{ref:gottschalk}.   

Within dimensional regularization the soft and collinear singularities 
in $(\Delta)H_i^{q,g}$ can be isolated using standard distribution-valued 
expansions \cite{ref:aem} of the type
\begin{equation}
\label{eq:ydist}
\yhat^\varepsilon (1-\yhat)^{-1+\varepsilon}=
\frac{1}{\varepsilon} \delta (1-\yhat)+\frac{1}{(1-\yhat)_+}+\varepsilon
\left\{\left[\frac{\ln(1-\yhat)}{1-\yhat}\right]_++\frac{\ln \yhat}{1-\yhat} 
\right\}\ \ \ ,
\end{equation}
where the `+'-distributions are defined in (\ref{eq:distdef}) in the Appendix. 
The relevant expansions for the initial state variable $\xi^\prime$ can be found in
Eqs.~(30)--(32) of \cite{ref:gottschalk}. After isolating soft and collinear 
poles and canceling the soft poles we can -- according to our discussion 
below Eq.~(\ref{eq:dps2}) -- transform the results to the final state charm quark
momentum scaling variable $\zeta \equiv p_c\cdot p_{s,g}/q\cdot p_{s,g}$ using
\begin{equation}
\label{eq:jaco}
\frac{d \yhat}{d \zeta} = \frac{s}{s-m_c^2}\ \ \ . 
\end{equation} 
Care has to be taken when applying the Jacobian in Eq.~(\ref{eq:jaco})
to distribution-valued expressions in the variable $\yhat$ as in Eq.~(\ref{eq:ydist}).
There the transformation does not merely amount to a multiplication with 
$d \yhat/ d \zeta$ but rather 
has to be found by changing the integration variable when folding the distribution 
with some test function such that:
\begin{equation}
\delta(1-\yhat)\rightarrow \delta(1-\zeta)\ \ \ ;
\ \ \ \frac{1}{(1-\yhat)_+}\rightarrow \frac{1}{(1-\zeta)_\oplus}\ \ \ ;
\ \ \ \left[ \frac{\ln(1-\yhat)}{1-\yhat}\right]_+\rightarrow 
\left[\frac{\frac{\ln(1-\zeta)}{1-\zeta_{\min}}}{1-\zeta}\right]_\oplus\ \ \ ,
\end{equation}
where the `$\oplus$'-distributions are defined again in (\ref{eq:distdef}),
$\lambda\equiv Q^2/(Q^2+m_c^2)$, and 
$\zeta_{\min}=(1-\lambda)\xi^\prime / (1-\lambda \xi^\prime)$.
Furthermore, because $\zeta_{\min}\rightarrow 1$ as $\xi^\prime \rightarrow 1$ 
\begin{equation}
\label{eq:deldel}
\delta(1-\xi^\prime) f(\xi^\prime,\zeta)=\delta(1-\xi^\prime)\delta(1-\zeta)
\left[ \int_{\zeta_{\min}}^1 d\alpha f(\xi^\prime,\alpha)\right]_{\xi^\prime=1}\ \ \ ,
\end{equation}
where $f$ may be either a function or distribution. We note here that by an 
analogous Jacobian transformation as in Eqs.~(\ref{eq:ydist})-(\ref{eq:deldel})
we could in principle obtain differential distributions in the transverse momentum
$p^T_c$ ($=\sqrt{s/4-m_c^2} \sin \theta^\ast$) of the produced charm quark as well. 

Using the standard tensor decomposition of the hadronic tensor,
the structure functions $F_{i=1,2,3}^{W^{\pm}}$ (unpolarized) and 
$g_{i=3,4,1}^{W^{\pm}}$ (polarized) refer to the following 
double [triple] differential CC $e^{\mp}p$ cross sections
\begin{eqnarray}
\label{eq:unpolxsec}
\frac{d^{2,[3]}\sigma^{e^{\mp}p}}{dx\ dy\ [dz]}\ &=&\ \frac{G_F^2 S_{ep}}
{2 \pi(1+Q^2/M_W^2)^2}\
\left[(1-y)F_2^{W^{\mp}}+y^2xF_1^{W^{\mp}}\pm
y(1-\frac{y}{2})xF_3^{W^{\mp}}\right]\\
\label{eq:polxsec}
\frac{d^{2,[3]} \Delta\sigma^{e^{\mp}p}}{dx\ dy\ [dz]}\ &=&\ \frac{G_F^2 S_{ep}}
{2 \pi(1+Q^2/M_W^2)^2}\
\left[(1-y)g_4^{W^{\mp}}+y^2xg_3^{W^{\mp}}\pm
y(1-\frac{y}{2})xg_1^{W^{\mp})}\right]
\end{eqnarray}
for a fully polarized lepton beam (left-handed $e^-$ or right-handed
$e^+$) scattering off an unpolarized (\ref{eq:unpolxsec}) 
or a polarized (\ref{eq:polxsec}) target and where $G_F$, $S_{ep}$ and $M_W$ denote the
Fermi coupling, available c.m.s.\ energy squared, and $W$-boson mass, respectively.
The longitudinally polarized cross section $d\Delta\sigma$ in
(\ref{eq:polxsec}) is defined as the difference
$(d\sigma^{\rightarrow}_{\Leftarrow} - d\sigma^{\rightarrow}_{\Rightarrow})$, where
$\Rightarrow$ ($\rightarrow$) denotes the direction of the proton (lepton) spin. 
In case the incident lepton is not fully polarized, i.e., $P_e\neq 1$, the r.h.s.\ of 
(\ref{eq:unpolxsec}) and (\ref{eq:polxsec}) have to be multiplied by an overall 
factor $(1+ P_e)/2$, which amounts to a factor $1/2$ for an unpolarized 
lepton beam $(P_e=0)$. One should note the change of sign of the $F_3$ and
$g_1$ terms in (\ref{eq:unpolxsec}) and (\ref{eq:polxsec}), respectively,
when the electron $e^-$ is replaced by a positron $e^+$. 
In Eqs.~(\ref{eq:unpolxsec}) and (\ref{eq:polxsec}) $x$ and $y$ are the standard 
kinematical deep inelastic variables (Bjorken scaling variable and inelasticity, 
respectively) and, moreover, $z\equiv p_D \cdot P / q \cdot P$ ($p_D$ being the 
charmed hadron's momentum) is a final state scaling variable.

The CC structure functions entering $d(\Delta)\sigma^{{e^-}p}$ in 
Eqs.~(\ref{eq:unpolxsec}) and (\ref{eq:polxsec}) are obtained by the following
convolutions\footnote{With obvious adjustments for $d \sigma^{e^+p}$ in 
Eqs.~(\ref{eq:inc}) and (\ref{eq:diff}) below:
$(\Delta) \bar{s} \rightarrow (\Delta ) s$ and 
$\{{\cal{F}}_3,{\cal{G}}_{3,4}\}\rightarrow - \{{\cal{F}}_3,{\cal{G}}_{3,4}\}$.}:
\begin{eqnarray}
\left\{{{\cal{F}}_i^c \atop {\cal{G}}_i^c} (x,Q^2) \right\} 
= \left\{{\bar{s} \atop \Delta \bar{s}} (\xi,\mu_F^2)\right\} &+&
\frac{\alpha_s(\mu_R^2)}{2\pi}
\left(\int_{\xi}^1 \frac{d\xi '}{\xi '} \left[
\left\{{H_i^q \atop \Delta H_i^q}  (\xi ',\mu_F^2,\lambda) \right\}
\left\{{\bar{s} \atop \Delta \bar{s}}(\frac{\xi}{\xi '},\mu_F^2) \right\}
\right. \right. \nonumber\\
&+& \left. \left. 
\left\{{H_i^g \atop \Delta H_i^g} (\xi ',\mu_F^2,\lambda) \right\}
\left\{{g\atop \Delta g}(\frac{\xi}{\xi '},\mu_F^2) \right\}
\right]\right)
\label{eq:inc}
\end{eqnarray}
\vspace*{-0.5cm}
\begin{eqnarray}  \nonumber
\left\{{{\cal{F}}_i^c \atop {\cal{G}}_i^c}(x,z,Q^2) \right\} 
&=& 
\left\{{\bar{s} \atop \Delta \bar{s}}(\xi,\mu^2_F) \right\}
\ D_c(z) \\ \nonumber &+&
\frac{\alpha_s(\mu^2_R)}{2\pi}
\int_{\xi}^1 \frac{d\xi '}{\xi '} 
\int_{\max(z,\zeta_{\min})}^1 \frac{d\zeta}{\zeta} 
\left[
\left\{{H_i^q\atop \Delta H_i^q}(\xi ',\zeta,\mu^2_F,\lambda) \right\}
\ \left\{{\bar{s}\atop \Delta \bar{s}}(\frac{\xi}{\xi '},\mu^2_F) \right\}
\right. \\ 
&+& \left.  
\left\{{H_i^g\atop \Delta H_i^g}(\xi ',\zeta,\mu^2_F,\lambda)\right\}
\ \left\{{g\atop \Delta g}(\frac{\xi}{\xi '},\mu^2_F)\right\}
\right] D_c(\frac{z}{\zeta}) \ \ . 
\label{eq:diff}
\end{eqnarray}
where
\begin{equation}
\label{eq:trans} 
\left\{{{\cal{F}}_1^c \atop {\cal{G}}_3^c}\right\} 
\equiv \left\{{F_1^c \atop {-g_3^c}}\right\}\ ;         
\ \ \ \left\{{{\cal{F}}_3^c \atop {\cal{G}}_1^c}\right\} 
\equiv \frac{1}{2}\left\{{{-F_3^c} \atop g_1^c}\right\}\ ; 
\ \ \ \left\{{{\cal{F}}_2^c \atop {\cal{G}}_4^c} \right\} \equiv 
\frac{1}{2\xi} \left\{ {F_2^c \atop {-g_4^c} } \right\}\ \ .
\end{equation} 
The superscript `$c$' in 
(\ref{eq:inc})-(\ref{eq:trans}) indicates that we restrict ourselves only to the  
charm production contribution to deep inelastic 
CC structure functions and the cross sections in (\ref{eq:unpolxsec}) and
(\ref{eq:polxsec}). For simplicity we 
set the factorization scale $\mu_F$ equal to the renormalization scale $\mu_R$ in 
(\ref{eq:inc}) and (\ref{eq:diff}) and fix both at $\mu_F^2=\mu_R^2\equiv Q^2+m_c^2$.
The coefficients $H_i^{q,g}$ can be found in \cite{ref:gkr1,ref:gkr2} while the 
polarized $\Delta H_i^{q,g}$ are new and listed in the Appendix. 
Please note that the $\Delta H_i^q(\xi,\zeta,\mu^2_F,\lambda)$ in the
Appendix (exactly as the $H_i^q(\xi,\zeta,\mu^2_F,\lambda)$ in 
\cite{ref:gkr2}) comprise terms of the type $[f(\xi)]_+ g(\xi)$ 
where $g$ is a singular function at $\xi=1$. These terms seem to be 
ill-defined at first sight. They are, however, completely well-behaved on the 
phase space of the double convolutions in Eq.~(\ref{eq:diff}) because the integration 
volume $\Delta \zeta =1-\zeta_{\min}$ vanishes at $\xi=1$ such that 
$[g(\xi)( 1-\zeta_{\min})]_{\xi=1}$ is always finite.

In Eq.~(\ref{eq:diff}) the (factorization scale independent) 
charm fragmentation function is taken as \cite{ref:peterson}
\begin{equation}
\label{eq:peterson}
D_c(z) = N \left\{ z \left[ 1-z^{-1}-\varepsilon_c/(1-z)
\right]^2\right\}^{-1}
\end{equation}
with the normalization constant $N$ being related to $\varepsilon_c$ via
$\int_0^1 dz D_c(z) = 1$. 
The `hardness' parameter will be fixed to be 
$\varepsilon_c = 0.06$ for our numerical applications in the
next section in agreement \cite{ref:gkr2,ref:kschie2} with fixed target 
neutrino data [4-6]. For our phenomenological considerations in the 
following section the precise value of $\varepsilon_c$ is, however, of minor importance.

An important comment about our final expressions for the $\Delta H_i^q$ 
given in (\ref{eq:h3qxz})-(\ref{eq:h1qxz}) should be made.
A naive calculation, as outlined above, gives:
\begin{equation}
\Delta {\tilde{H}}_{i=3,4,1}^q(\xi,\zeta,\mu^2_F,\lambda) 
= H_{i=1,2,3}^q(\xi,\zeta,\mu^2_F,\lambda)-
4 C_F (1-\xi)\delta(1-\zeta)\ \ \ .
\end{equation}
The difference $4 C_F (1-\xi)\delta(1-\zeta)$ stems, however, from a too naive  
factorization of the initial state collinear $s\rightarrow s g$ singularity
in the polarized case in $n\neq 4 $ dimensions. In order to restore helicity 
conservation at the strange quark-gluon vertex \cite{ref:vogelsang}
the finite renormalization
\begin{equation}
\Delta {\tilde{H}}_{i=3,4,1}^q(\xi,\zeta,\mu^2_F,\lambda) 
\rightarrow \Delta {\tilde{H}}_{i=3,4,1}^q(\xi,\zeta,\mu^2_F,\lambda)
+4 C_F (1-\xi)\delta(1-\zeta) 
= H_{i=1,2,3}^q(\xi,\zeta,\mu^2_F,\lambda)
\end{equation} 
has to be considered, leading to our final results in the Appendix.
These NLO ($\overline{\mathrm{MS}})$ coefficient functions
$\Delta H^q_{3,4,1}(\xi,\zeta,\mu_F^2,\lambda)$ hence coincide with the corresponding 
unpolarized expressions $H^q_{1,2,3}(\xi,\zeta,\mu_F^2,\lambda)$ \cite{ref:gkr2}, 
as must be due to the same tensorial structure at the partonic level,
the nature of the V-A interactions, and helicity conservation
at the (massless) strange quark-gluon vertex.
It should be remarked that the fully 
inclusive quark coefficients $\Delta H_i^q(\xi,\mu^2,\lambda)$ have been
already obtained in \cite{ref:svw}, and we fully agree with these results.

For the $H_i^g$  the relation between the small 
strange quark mass limit of the fully massive ($m_{s,c}\neq 0$) BGF process and the 
$\overline{\mathrm{MS}}$ results of \cite{ref:gkr2} has been established in 
Eq.~(5) of Ref.~\cite{ref:kschie3}. We note here that analogously the 
$\Delta H_i^g(\xi,\zeta,\mu_F^2,\lambda)$ in (\ref{eq:higxz}) can also be obtained from 
the 
$m_s\rightarrow 0$ limit of the general, fully massive spin-dependent BGF
expressions in Eq.~(10) of \cite{ref:schienbein}. This is a non-trivial 
cross check of our \msbar\ results. Similarly our expressions for 
$\Delta H_i^g(\xi,\mu_F^2,\lambda)$ in (\ref{eq:higx}) agree with the $k_{T}^{\min}$ 
(transverse momentum cut-off) regulated results in Eq.~(13) of \cite{ref:vw} for 
$k_{T}^{\min}=0$ in the limit $m_s\rightarrow 0$. 
Finally, taking the limit $m_c\rightarrow 0$ of our 
$\Delta H_i^{q,g}(\xi,,\mu_F^2,\lambda)$ we recover -- apart from obvious 
collinear logs -- the massless results in \cite{ref:deflorian} after transforming 
them to the conventional \msbar\ scheme by setting the 
$\widetilde{\Delta f_i^{q,g}}=0$ in \cite{ref:deflorian}.
The results in \cite{ref:deflorian} originally refer to a scheme
with a non-vanishing gluonic contribution to the first moment
of $g_1$, $g_3$ and $g_4$ as do the $k_{T}^{\min}$ regulated
results in Eq.~(15) in \cite{ref:vw} when taking the massless limit.  

\section{Determining $\bar{s}$ and $\Delta \bar{s}$ at HERA}
Equipped with the required technical framework, we now turn to
a detailed discussion and numerical analysis of the prospects of 
determining $\bar{s}$ and $\Delta \bar{s}$ at HERA in the future.

To begin with, Fig.~2 shows the unpolarized $z$ differential $\bar{D}$ meson 
production cross section $e^-p\rightarrow \bar{D} X$ according to 
Eq.~(\ref{eq:unpolxsec}) for a fixed value of $x=0.05$ and where we have 
integrated over $Q^2=S_{ep} x y>500\,\mathrm{GeV}^2$ employing the cut 
$0.01\le y \le 0.9$, 
with $\sqrt{S_{ep}}=300\,\mathrm{GeV}$ for HERA.
Unless otherwise stated we use the GRV-94 \cite{ref:grv94} distributions in all our 
unpolarized calculations\footnote{\label{footnr}This is mainly to avoid any 
inconsistencies due to different $\Lambda_{QCD}$ values, spurious
violations of the positivity requirement $|\Delta f|\le f$, etc., in the 
calculation of the polarization asymmetries (\ref{eq:asym}) below, since the
spin-dependent GRSV parton distributions \cite{ref:grsv}, which we adopt, 
were being set-up with the unpolarized GRV-94 distributions \cite{ref:grv94} as 
reference. 
Anyway, the results obtained using the recent GRV update \cite{ref:grv98} are
almost indistinguishable.}. The different contributions to the NLO
cross section $d\sigma/dx dz$ are shown separately in Fig.~2 to demonstrate the 
impact of the genuine NLO gluon induced subprocess.
Needless to say, 
only the total NLO cross section is a physically meaningful observable.
As can be seen, the gluonic contribution becomes increasingly important
and eventually dominates at small values of $z$, where the Born approximation becomes 
completely meaningless. The sharp rise for $z\rightarrow 0$ can be traced back to 
the $1/\zeta$ behaviour in the $g\rightarrow c\bar{c}$ splitting in the $u$-channel 
subprocess depicted on the r.h.s.\ of Fig.~1(d). 
This behaviour should {\it not} be considered as a destabilization of the perturbative 
expansion by NLO corrections because it is entirely due
to the {\it first} contribution from charm quarks produced by strong interaction 
dynamics which one even should {\it expect} to be important at high $Q^2$.
Note that in the region of BGF dominance around $z \lesssim 0.1$ the production 
dynamics and the resulting steep shape is very similar to the neutral current case 
\cite{ref:kschie2}, where the gluon fusion production channel is 
-- within fixed order perturbation theory -- classified as {\it leading order}. 
To avoid any confusion we will, nevertheless, in the following continue to 
count all \oalps\ contributions as `NLO' in the CC case under consideration here. 
Let us note that, on top of an $\bar{s}(x)$ measurement, the mere observation of 
a rising cross section towards small $z$ at HERA would be an interesting experimental
confirmation of the underlying QCD dynamics and evolution effects because such a 
behaviour is completely absent at fixed target scales [18,4-7], where the 
contribution from the charm quark in the u-channel of the BGF is small. 
Integrated over $z$, the steep small $z$ rise results in a quasi-collinear 
logarithm $\sim \ln Q^2/m_c^2$, which may lead to substantial gluonic corrections 
in the inclusive cross section $d\sigma /dx$ unless they are removed by acceptance 
losses or suitable cuts, as will be discussed in more detail below.

Both (gluon- {\it and} quark-initiated) NLO corrections are sizeable in the 
entire $z$ range as can be inferred from comparing the Born result
with the individual NLO contributions
in Fig.~2. The NLO corrections tend to soften the $\bar{D}$ meson
$z$ spectrum and shift the peak due to the Peterson fragmentation 
spectrum (\ref{eq:peterson}), located at around $z\simeq 0.75-0.8$ 
in the Born term, towards smaller values of $z$.
However, we have used NLO distributions and a NLO value for 
$\varepsilon_c$ in (\ref{eq:peterson}) for the calculation of the Born term, 
and hence a LO and NLO comparison does not reflect a real $K$ factor, i.e., 
the ratio $d\sigma_{NLO}/d\sigma_{LO}$, which has to be obtained
in a consistent LO and NLO calculation using LO and NLO densities
and parameters, respectively. We refrained from doing so here, since
on the one hand possible differences between the LO and NLO strange densities 
are hardly known yet, and on the other hand our main purpose is to illustrate 
the impact of the NLO corrections undiluted by different choices of the parton 
densities and other parameters. Furthermore, $K$ factor considerations are 
somewhat misleading for the differential observable considered here 
because the `LO' Born term is at the partonic level by construction\footnote{It 
sums up the leading {\it collinear} logs.} sharply peaked in the forward (proton) 
direction $\sim \delta(1-\zeta )\propto \delta(1+\cos \theta^\ast)$,  
and a continuous spectrum in $z$ is only achieved by smearing
the delta peak with the Peterson function in (\ref{eq:peterson}), i.e., purely by 
non-perturbative hadronization effects. Only in NLO a differential distribution 
which covers the full phase space is obtained already at the (perturbative) partonic 
level and then translates into a realistic hadronic ${\bar D}$ meson 
momentum spectrum.
It should be mentioned as well that in the very large $z$ region, $z\gtrsim 0.85$, 
perturbatively large $\ln(1-z)$ terms in the NLO subprocesses have to be 
resummed \cite{ref:melenason}, and non-perturbative higher twist contributions 
will perhaps become relevant \cite{ref:nasonwebber} such that our results in 
this region cannot be trusted (as is obvious, of course, from the negative 
cross section obtained here).

In Fig.~3 we compare the results for the $z$ integrated $\bar{D}$ meson 
production cross section $d\sigma/dx$ for two cases: either integrated 
over the entire $z$ range (inclusive cross section) or 
restricted to the region $0.2<z<1$, which turns out to be a 
suitable theoretical cut to strongly reduce the gluonic `background' and
perhaps also mimics a poorer experimental resolution a low $z$ \cite{ref:bgnpz}.
As expected from our results in Fig.~2 we obtain again large NLO 
corrections in the first case, mainly due to the sharp rise 
for $z\rightarrow 0$ in NLO, whereas in the second case the corrections
cancel to a large extent. This cancellation can be readily understood
from the results shown in Fig.~2: both NLO contributions change sign
at around $z\simeq 0.5-0.7$ and roughly integrate to zero in
the chosen $z$ range\footnote{One can restrict the integrations 
also to the range $0.2<z\lesssim0.85$ 
to avoid the region in the cross section where $\ln(1-z)$ logarithms become
dominant. This would lead only to a slightly larger NLO result.}
$0.2<z<1$. The small-$z$ region where the gluonic correction becomes 
dominant is simply left out, thereby minimizing the residual NLO effects.
It should be mentioned here that in the charge conjugated case
$e^+p\rightarrow D X$ the $d_v \rightarrow c$ valence enhancement
at large $x$, which we have briefly discussed in the Introduction, becomes 
significant at the $\sim 10\%$ level around $x\sim 0.1$, but dominant
only for very large $x \gtrsim 0.5$.     

A comment should be made, however, about the NLO corrections which seem to be 
anomalously large at small values of $x$ in the fully inclusive case  
($0<z<1$), where they are dominated by a large quasi-collinear 
logarithm $\ln Q^2/m_c^2$.
One may attribute, for the moment, this logarithm to a corresponding
$W^- c\rightarrow s$ contribution\footnote{Obviously such a 
process would not be considered as CC charm production since there is no charm 
activity in the final state (see also the discussion below).} by introducing 
a resummed, massless charm density $c(x)$. 
The large NLO corrections seem then in turn to imply a larger contribution from 
charm than strange quarks, i.e., $c(x)>s(x)$ in such a language, see also Fig.~2 
(r.h.s.) in \cite{ref:barone}, which would be theoretically not very appealing 
and would therefore seriously question the perturbative reliability of the fixed order 
logarithm $\ln Q^2/m_c^2$. One has to be careful with such interpretations 
though, since for small values of $x$ and hence large values of $y$ of about 0.7 at 
HERA, the structure function $F_3$ becomes important in (\ref{eq:unpolxsec}). 
In $F_3^{ep}$ $s$ and $c$ enter with negative and positive signs, respectively. 
Taking into account the $x$ and $y$ dependent weights in front of $F_1$, $F_2$, 
and $F_3$ in (\ref{eq:unpolxsec},) it turns out that the negative $s$ contribution 
in $F_3$ cancels to a considerable amount the positive contributions in $F_1$ and $F_2$, 
whereas all charm contributions add up. Therefore the surprising result that 
charm quarks contribute a larger portion to $d \sigma$ than strange quarks 
(mimicing `$c(x)>s(x)$') is simply an effect of the electroweak couplings in 
that particular kinematic region.

Another indication that the large gluonic correction $\sim \ln Q^2/m_c^2$
in Fig.~3 does not necessarily call for collinear resummations comes from
an ${\cal{O}}(\alpha_s^2)$ calculation of CC heavy quark production
in the asymptotic limit $Q^2\gg m_c^2$ \cite{ref:buza}. Here it was
found that although the \oalps\ gluonic contribution is large, 
the ${\cal{O}}(\alpha_s^2)$ terms hint at a completely stable and well-behaved 
fixed order perturbation series. As an indication of the perturbative reliability 
of our results it should be  furthermore noted that the dependence of our results 
in Figs.~2 and 3 on the precise value of the factorization scale $\mu_F$ is rather 
weak. Variations of $\mu_F^2$ in the range 
$0.1(Q^2+m_c^2)\le \mu_F^2\le 10(Q^2+m_c^2)$, i.e., by two orders of magnitude 
around our default value $\mu_F^2=Q^2+m_c^2$, change the results for 
$d\sigma/dx dz$ and $d\sigma/dx$ shown in Figs.~2 and 3 by at most $\sim \pm 10\%$.
On top of this comes a more practical reason for our preference for fixed order 
perturbation theory in the CC case. As mentioned above the resummation of 
quasi-collinear logs $\sim \ln Q^2/m_c^2$ requires simultaneously the 
introduction of a masslessly evolved charm density $c(x)$ entering a 
$W^- c\rightarrow s$ production channel, where the corresponding ${\bar{c}}$ 
from the gluon splitting must be thought of as hiding in the hadronic debris. 
This means that we loose any information whether the event will be tagged as a 
charm event experimentally. Even if we assume that the ${\bar c}$ escapes 
always {\it un}detected, we have no possibility for a 
gauge and renormalization group invariant separation
of this production channel from the (tagged) rest of the cross section. 
On the other hand, in a differential fixed order calculation we can, 
as demonstrated above, apply suitable $z\rightarrow 0$ (or $p_c^T \rightarrow 0$)
cuts to exclude the quasi-collinear region. 

It should be mentioned that the $x$ value in Fig.~2 has been chosen to guarantee 
a sizeable cross section, i.e., a realistic  chance to actually measure $\bar{s}$. 
For larger values of $x$, $d\sigma/dx dz$ shows qualitatively similar features as 
the ones illustrated in Fig.~2 but 
at a much reduced cross section as can be inferred from Fig.~3. Of course, 
the gluon contribution becomes less important with increasing $x$, and the rise 
for $z\rightarrow 0$  becomes much less pronounced. 
With the upcoming luminosity upgrade at HERA it should be feasible to study 
these CC reactions with sufficient statistics.
The obtained cross sections in Figs.~2 and 3 are all in the ball-park 
of about $30-50\,\mathrm{pb}$ for $x\simeq 0.05$ and thus seem to be measurable, 
assuming an integrated luminosity of ${\cal{L}}=200-500\,\mathrm{pb}^{-1}$ in the 
future, even if one takes a rather small charm detection efficiency of about 
$1-2\%$ into account. The charm detection is actually the limiting factor for 
such measurements, in particular in the polarized case as will be shown below.
The $z$-integrated cross section including a cut $z\gtrsim 0.2$ in Fig.~3 is 
particularly suited for a measurement of the strange density at HERA since the 
`background' from BGF drops out almost completely, and the full NLO
cross section can be approximated by its Born term, which is directly sensitive 
to $\bar{s}$. However, within the achievable accuracy it seems to be virtually
impossible to distinguish between different currently available strange densities as 
is illustrated in Fig.~4. The differences between the results obtained using the 
GRV-94 densities \cite{ref:grv94} (or the recent update \cite{ref:grv98}) or 
the MRST distributions \cite{ref:mrst} are not very pronounced
at $Q^2\ge 500\,\mathrm{GeV}^2$ since they are washed out by the $Q^2$ evolution. 
A possible improvement would be to add the results for $D$ and $\bar{D}$ 
meson production, i.e., the results obtained for $e^-p$ and $e^+p$ CC reactions, 
which would obviously double the statistics. The price to pay is of course that 
one would loose any sensitivity to possible differences between $s$ and $\bar{s}$ 
\cite{ref:brodsky}. However, it seems to be difficult anyway to test the latter 
at HERA unless $s$ and $\bar{s}$ would drastically differ which seems to be not 
very realistic. To finish this discussion it should be 
stressed that the differences between the GRV and MRST results in Fig.~4 are 
indeed mainly due to the different assumptions about the strange density.
Any differences in the gluon distribution are not important in that particular
kinematic region (in Fig.~4(b) the gluon contribution almost cancels anyway), for 
instance, using the `large' gluon or `small' gluon set of MRST, MRST(g$\uparrow$) 
and MRST(g$\downarrow$) \cite{ref:mrst}, respectively, instead, hardly leads to any 
changes.

Let us now turn to the polarized case. As already mentioned in the Introduction, 
such measurements could be performed at HERA as well provided that the option to 
longitudinally polarize both beams \cite{ref:herapol} will be realized in the future. 
Since large polarized targets as required for neutrino DIS are not likely to be ever 
build, it should be stressed that HERA would be a unique place to study CC DIS 
(among other reactions which can be only analyzed at a polarized $ep$ collider, 
see \cite{ref:herapol}).

Figs.~5 and 6 show the polarized $z$ differential and integrated $\bar{D}$ meson 
production cross sections according to Eq.~(\ref{eq:polxsec}) in a similar way as 
above in the unpolarized case in Figs.~2 and 3, respectively. Unless otherwise 
stated we use the GRSV `standard' set of spin-dependent parton densities 
\cite{ref:grsv} in our calculations, which provides a rather large, negatively 
polarized strange density, i.e., $\Delta s(x,Q^2)=\Delta \bar{s}(x,Q^2)<0$ 
for all values of $x$ and $Q^2$. 
For the differential cross section $d\Delta \sigma/dxdz$ in Fig.~5 
one observes the same qualitative feature as in the unpolarized case: 
the NLO corrections become increasingly important for small values of $z$
due to sharp rise of the gluonic contribution. However, the polarized gluon density 
$\Delta g$ is only very weakly constrained by presently available data, and hence 
the actual size of the gluonic correction is extremely uncertain and strongly 
dependent on the chosen set of parton densities. A meaningful measurement
of $\Delta \bar{s}$ is therefore only possible if $\Delta g$ becomes better 
constrained in the future, which is not unlikely in view of the forthcoming
experiments at BNL-RHIC and CERN (COMPASS), or -- even better -- if the gluonic 
`background' can be eliminated or largely reduced by some suitable cuts. 
The latter can be partly achieved for the $z$ integrated $\bar{D}$-meson cross section 
$d\Delta \sigma/dx$ by introducing 
-- as in Fig.~3 above -- a lower cut-off $z\gtrsim 0.2$ for the integration
as can be inferred  from Fig.~6. Contrary to the unpolarized case, the gluonic 
correction does not fully drop out here, since it is not oscillating in the region
$0.2<z<1$ (at least not for the chosen $\Delta g$). However, this observable 
nevertheless seems to be best suited for a determination of the strange sea, 
but some knowledge of $\Delta g$ would be certainly desirable to disentangle 
its `background' more precisely. It should be mentioned that the dependence on 
the factorization scale is rather weak, similar to our observations for the unpolarized
case.

Finally, let us study the sensitivity of CC charm production to the unknown 
spin-dependent strange density by comparing the results obtained for different, 
extreme choices of $\Delta \bar{s}$. First of all it should be noted that the 
actual observable is the spin asymmetry for $D$ meson production
\begin{equation}
\label{eq:asym}  
A^D \equiv \frac{\int_{z_{\min}}^1 dz\; d\Delta \sigma/dx/dz}
{\int_{z_{\min}}^1 dz \;d\sigma/dx/dz} 
\end{equation}
rather than the polarized cross sections for $z_{\min}=0$ and $0.2$ 
shown in Fig.~6. $A^D$ in (\ref{eq:asym}) is simply related to a measurement of
the counting rate asymmetry for parallel and anti-parallel alignment of the proton 
and lepton spins and does not require the determination of the absolute normalization. 
Furthermore, other experimental uncertainties conveniently drop out in 
the ratio (\ref{eq:asym}). 

Fig.~7 shows our results for $d\Delta \sigma/dx$ and $A^D$ in longitudinally 
polarized $e^-p$ and $e^+p$ collisions for $z_{\min}=0$ and $0.2$.
Apart from the GRSV `standard' set, we now adopt also the GRSV `valence' set 
\cite{ref:grsv}, which, on the contrary, has a small positive strange density 
in the relevant $x$ region whereas the gluon distribution is practically unchanged.
While the cut $z>0.2$ merely changes the size of the cross section and hardly 
effects the asymmetry $A^D$ for the `standard' set, the influence of the cut on 
$A^D$ is more pronounced for the `valence' set as can be inferred from comparing 
Figs.~7(b) and (d). The oscillating behaviour in the `valence' 
case stems from the interplay of Born and NLO contributions with different signs. 
Also shown in Fig.~7 is the expected statistical accuracy $\delta A^D$ for such a 
measurement at a polarized HERA 
\begin{equation}
\label{eq:error}
\delta A^D = \frac{1}{P_p} \frac{\sqrt{1-P_p^2 {A^D}^2}}
           {\sqrt{{\cal{L}} \int dx\, d\sigma/dx 
           \,\varepsilon_{eff}^c (1+P_e)/2}}
\end{equation}
assuming an integrated luminosity of ${\cal{L}}=500\,\mathrm{pb}^{-1}$,
$70\%$ beam polarizations $P_p$ and $P_e$, and where we have integrated
over three bins in $x$. As can be seen the results for either $e^-p$ or $e^+p$ 
collisions for the two different sets of parton densities can be distinguished 
within the error bars and hence some information on $\Delta s$ can be extracted. 
However, there is a severe catch: In Fig.~7 
a charm detection efficiency of $100\%$ was assumed, i.e., 
$\varepsilon_{eff}^c=1$, as was also used in previous LO calculations \cite{ref:maul}, 
but which is completely unrealistic. With present-day values for 
$\varepsilon_{eff}^c$ at HERA of much less than $5\%$ a measurement of $\Delta s$ 
via CC $D$ meson production is certainly impossible.
However, until a polarized HERA could be realized in the future, some 
progress on charm detection might be possible.
Furthermore, given the possibility that runs will be made with $e^-$ and
$e^+$ beams these results could be added to double the statistics.
Of course, possible changes of sign in $d\Delta \sigma$ have to taken into
account by taking the absolute value, and it would be 
somewhat less clear how to extract $\Delta \bar{s}$ in such a case.

\section{Summary}
We have presented the complete NLO QCD framework for CC mediated
inclusive charm and momentum $z$ differential $D$ meson production in
DIS with unpolarized and longitudinally polarized beams.
The calculated \oalps\ coefficient functions refer to the \msbar\
fixed flavor scheme which fully takes into account the mass of
the produced heavy (charm) quark. Our unpolarized results fully agree
with previous calculations while the spin-dependent expressions are
entirely new. Special emphasis was put on technical subtleties 
in the $z$ differential case and due to the appearance of $\gamma_5$ in
$n$ dimensional regularization.

Exploiting our analytical results we have performed a detailed 
phenomenological analysis of the prospects of determining the unpolarized
and polarized strange density from CC $D$ meson events at HERA.
It was shown how to reduce the `unwanted' contribution from the
genuinely NLO gluon initiated subprocess considerably by imposing
a lower cut-off on the $D$ meson momentum fraction $z$ for $z$
integrated rates. Furthermore, it was argued that 
the sizeable NLO corrections observed for inclusive charm
and momentum $z$ differential $D$ meson production  
can be both understood and even expected in the kinematical domain
of HERA due to the peculiarities arising from the mixture of weak and strong
interactions in the case of CC charm production.

It turned out that the rather small charm detection efficiency is the
main limiting factor in pinning down the strange sea at HERA. 
Nevertheless, the {\em unpolarized} strange density was shown to be 
measurable with sufficient accuracy (a decent luminosity permitting),
and HERA can hopefully contribute to a better understanding of the
so far only weakly constrained strange distribution in the future.
Unfortunately, in the polarized case, where much less is known about the
flavor decomposition of the sea, a useful measurement of $\Delta \bar{s}$
from the CC $D$ meson spin asymmetry cannot be performed without a 
significantly improved charm detection efficiency.

\section*{Acknowledgements}
The work of S.K.\ has been supported in part by the 
`Bundesministerium f\"{u}r Bildung, Wissenschaft, Forschung und
Technologie', Bonn. 

\newpage
\setcounter{equation}{0}
\def\theequation{A\arabic{equation}}
\section*{Appendix}
Here we list the expressions for the NLO (\msbar) CC coefficient functions
$\Delta H_{i=3,4,1}^{q,g}$ for heavy quark (charm) production appearing in
Eqs.~(\ref{eq:inc}) and (\ref{eq:diff}).
The $\zeta$ differential fermionic NLO (\msbar) coefficients 
$\Delta H_{i=3,4,1}^q(\xi,\zeta,\mu_F^2,\lambda)$ in (\ref{eq:diff}),
as obtained from calculating the the subprocess $W^+ s \rightarrow g c$ and 
the virtual corrections to $W^+ s \rightarrow c$, coincide with the 
unpolarized functions $H_{i=1,2,3}^q(\xi,\zeta,\mu_F^2,\lambda)$ in 
\cite{ref:gkr2}. They will be nevertheless also given here for completeness:     
\begin{eqnarray} \nonumber
\Delta H_3^q(\xi,\zeta,\mu^2_F,\lambda) &=&
\delta (1-\zeta )\  \left\{\  \Delta P_{qq}^{(0)}(\xi)  \ln 
\frac{Q^2+m_c^2}{\mu^2_F}  \right.  \\ \nonumber &+& \left.
\frac{4}{3} \left[ 1-\xi + (1-\xi) \ln \frac{(1-\xi)^2}{\xi(1-\lambda
\xi)} - 2 \xi \frac{\ln \xi}{1-\xi} + 2 \xi 
\left( \frac{1}{1-\xi} \ln \frac{(1-\xi)^2}{1-\lambda \xi} \right)_+
\right] \right\} \\ \nonumber
&+& \frac{4}{3} \left\{ - \delta(1-\xi) \delta(1-\zeta) \left[
\frac{1}{2} \left( \frac{1+3 \lambda}{\lambda}\ K_A
+\frac{1}{\lambda} \right)\ +\ 4\ +\ 
\frac{\pi^2}{3} \right] \right. \\ \nonumber
&+& \frac{1-\xi}{(1-\zeta)_{\oplus}}\ +\ (1-\zeta)\left(
\frac{1-\lambda \xi}{1-\xi}\right)^2 \left[ \frac{1-\xi}{(1-\lambda\xi)^2}
\right]_+ \\ \nonumber
&+&  2\ \frac{\xi}{(1-\xi)_+}\ \frac{1}{(1-\zeta)_{\oplus}}
\left[ 1-(1-\zeta)\ \frac{1-\lambda \xi}{1-\xi} \right] \\
\label{eq:h3qxz}
&+& \left. 2\ \xi \left[ 1-(1-\zeta)\ \frac{1-\lambda \xi}{1-\xi}
\right] \right\} \\  \nonumber
\Delta H_4^q(\xi,\zeta,\mu^2_F,\lambda) &=& 
\Delta H_3^q(\xi,\zeta,\mu^2_F,\lambda)
+ \frac{4}{3} {\Bigg\{} \delta(1-\xi) \delta(1-\zeta) K_A  \\
&-& \left. 2 \left(\xi(1-3\lambda)[1-(1-\zeta)\frac{1-\lambda \xi}{1-\xi}]
+(1-\lambda)\right)\right\} \\
\Delta H_1^q(\xi,\zeta,\mu^2_F,\lambda) &=& 
\Delta H_3^q(\xi,\zeta,\mu^2_F,\lambda)
\ + 2\ \frac{4}{3}\left\{(1-\xi)[1-(1-\zeta)\frac{1-\lambda\xi}{1-\xi}]
-(1-\lambda\xi)\right\} 
\label{eq:h1qxz}
\end{eqnarray}
where we have defined $\lambda\equiv Q^2/(Q^2+m_c^2)$ and  
$K_A\equiv\frac{1}{\lambda}(1-\lambda)\ln(1-\lambda)$ and
$\Delta P_{qq}^{(0)}(\xi)=\frac{4}{3}\left(\frac{1+\xi^2}{1-\xi}\right)_+$
denotes the LO $q\rightarrow q$ splitting function.
The `+' and `$\oplus$' distributions in (\ref{eq:h3qxz})-(\ref{eq:h1qxz}) are defined 
by
\begin{equation}
\label{eq:distdef}
\int_0^1 d\xi\ \frac{f(\xi)}{(1-\xi)_+}=
\int_0^1 d\xi\ \frac{f(\xi)-f(1)}{1-\xi}\ ,\ \ \
\int_{\zeta_{\min}}^1 d\zeta\ \frac{f(\zeta)}{(1-\zeta)_{\oplus}}=
\int_{\zeta_{\min}}^1 d\zeta\ \frac{f(\zeta)-f(1)}{1-\zeta}
\end{equation}
with $\zeta_{\min}= (1-\lambda)\xi / (1-\lambda \xi)$. 

When integrated over $\zeta$, the results
given in (\ref{eq:h3qxz})-(\ref{eq:h1qxz}) reduce to the inclusive coefficients 
\begin{equation}
\Delta H_i^q(\xi,\mu^2_F,\lambda) \equiv \int_{\zeta_{\min}}^1 d\zeta
\ \Delta H_i^q(\xi,\zeta,\mu^2_F,\lambda)
\end{equation}
appearing in (\ref{eq:inc}), where
\begin{equation}
\Delta H_i^q(\xi,\mu^2_F,\lambda)\ =\ \left[\Delta P_{qq}^{(0)}(\xi)\
\ln\frac{Q^2+m_c^2}{\mu^2_F}\ +\ \Delta h_i^q(\xi,\lambda)\right]
\end{equation}
with
\begin{eqnarray}
\label{eq:smallhq}
\Delta h_i^q(\xi,\lambda)\ =\ \frac{4}{3} &\bigg\{& h^q+A_i\ \delta
(1-\xi)+B_{1,i}\ \frac{1}{(1-\xi)_+} \nonumber\\
 &+& \left. B_{2,i}\ \frac{1}{(1-\lambda
\xi)_+}+B_{3,i}\ \left[\frac{1-\xi}{(1-\lambda \xi)^2}\right]_+\right\}
\end{eqnarray}
and
\begin{eqnarray}  \nonumber
h^q\ =\ &-&\left(4+\frac{1}{2\lambda}+\frac{\pi^2}{3}+\frac{1+3\lambda}
{2\lambda}\ K_A\right)\delta(1-\xi) \\
 &-& \frac{(1+\xi^2)\ln \xi}
{1-\xi}+(1+\xi^2)\left[\frac{2\ln (1-\xi)-\ln (1-\lambda \xi)}{1-\xi}\right]_+
\ \ \ .
\end{eqnarray}
The coefficients in (A7) for $i=3,4,1$ are given in Table 1 and agree with
the results presented in 
\cite{ref:svw,ref:gkr1}.

\noindent
Table 1. Coefficients for the expansion of $\Delta h_i^q$ in (\ref{eq:smallhq}).\\
\begin{tabular*}{\textwidth}{@{~}l@{\extracolsep\fill}llll}
\hline\hline
$i$ & $A_i$ & $B_{1,i}$ & $B_{2,i}$ & $B_{3,i}$ \\  \hline
$3$ & $0$ & $1-4\xi+\xi^2$ & $\xi-\xi^2$ & $\frac{1}{2}$ \\
$4$ & $K_A$ & $2-2\xi^2-\frac{2}{\xi}$ & $\frac{2}{\xi}-1-\xi$ & $\frac{1}{2}$ \\
$1$ & $0$ & $-1-\xi^2$ & $1-\xi$ & $\frac{1}{2}$ \\ \hline\hline \\
\end{tabular*}

The $\zeta$ differential NLO (\msbar) gluonic coefficient functions 
$\Delta H_i^g(\xi,\zeta,\mu_F^2,\lambda)$ for heavy quark (charm) production 
in (\ref{eq:diff}), as calculated from the BGF subprocess 
$W^+ g \rightarrow c {\bar{s}}$, are given by
\begin{eqnarray} \nonumber
\Delta H^g_{i={3,4 \atop 1}}(\xi,\zeta,\mu^2_F,\lambda) &=&
\delta(1-\zeta) \left\{
\Delta P_{qg}^{(0)}(\xi)\left[\ln\frac{Q^2+m_c^2}{\mu^2_F} +
\ln\frac{(1-\xi)^2}{\xi(1-\lambda \xi)}\right]+(1-\xi)\right\} \\
&+& \left[\frac{1}{(1-\zeta)_{\oplus}}\mp \frac{1}{\zeta}\right]
\Delta P_{qg}^{(0)}(\xi)\ +\ \Delta h_i^g(\xi,\zeta,\lambda)
\label{eq:higxz}
\end{eqnarray}
where
\begin{eqnarray}  \nonumber
\Delta h_3^g(\xi,\zeta,\lambda) &=&
-\xi (1-\lambda ) \left[ \frac{1}{\zeta^2}-\frac{2}{\zeta} \right] \\ \nonumber
\Delta h_4^g(\xi,\zeta,\lambda) &=&
\frac{1-\lambda}{\zeta^2} \xi (1-2 \lambda) + \frac{1}{\zeta} \left[
2 \xi (1-\lambda^2) - (1-\lambda) \right]  \\ \nonumber
\Delta h_1^g(\xi,\zeta,\lambda) &=&
\xi (1-\lambda ) \left[ \frac{1}{\zeta^2} - \frac{2}{\zeta} \right]
+1-2 \lambda \xi 
\end{eqnarray}
with $\Delta P_{qg}^{(0)}(\xi)=\frac{1}{2}[2\xi-1]$ denoting the LO
$q\rightarrow g$ splitting function.
The $\oplus$ distribution is already defined in (\ref{eq:distdef}). 
When integrated over $\zeta$, the results in (\ref{eq:higxz})
reduce to the inclusive coefficients in (\ref{eq:inc}), i.e.,
\begin{equation}
\Delta H_i^g(\xi,\mu^2_F,\lambda) \equiv \int_{\zeta_{\min}}^1 d\zeta
\ \Delta H_i^g(\xi,\zeta,\mu^2_F,\lambda)\ \ \ ,
\end{equation}
where
\begin{equation}
\label{eq:higx}
\Delta H^g_{i={3,4 \atop 1}}(\xi,\mu^2_F,\lambda)\ =\
\left[\Delta P_{qg}^{(0)}(\xi)\left(\mp L_{\lambda}+
\ln\frac{Q^2+m_c^2}{\mu^2_F}+\ln\frac{(1-\xi)^2}{\xi(1-\lambda\xi)}
\right)+\Delta h_i^g(\xi,\lambda)\right]
\end{equation}
with 
\begin{eqnarray*}
L_{\lambda}\equiv \ln \frac{1-\lambda \xi}{(1-\lambda )\xi}
\end{eqnarray*}
and
\begin{eqnarray}  \nonumber
\Delta h_3^g(\xi,\lambda) &=&
2 (1-\lambda)\  L_{\lambda}\ \xi \\ \nonumber
\Delta h_4^g(\xi,\lambda) &=&
2 (1-\lambda)\ (1-\xi)+(1-\lambda)\ L_{\lambda}\ [2\xi(1+\lambda)-1] 
\\ \nonumber
\Delta h_1^g(\xi,\lambda) &=&
(1-\xi)\ [ 4-1/(1-\lambda\xi) ]
-2(1-\lambda)\ L_{\lambda}\ \xi\ \ \ \ .
\end{eqnarray}
\newpage
%


\newpage
\section*{Figure Captions}
\begin{description}
\item[Fig.~1] Feynman diagrams contributing to the CC 
massive charm ($p_c^2=m_c^2$) production 
up to ${\cal{O}}(\alpha_s)$: Born term (a), real gluon emission (b),
virtual corrections (c), and boson gluon fusion (d).
The Cabibbo suppressed contributions are obtained by substituting all 
$s$ by $d$ quarks.
The relevant diagrams for $\bar{s}\rightarrow \bar{c}$ transitions are 
obtained by reversing all quark lines in (a)-(d).
\item[Fig.~2] $z$ differential CC $\bar{D}$ meson production cross section
$e^-p\rightarrow \bar{D} X$ in NLO obtained by integrating Eq.~(\ref{eq:unpolxsec}) 
over the range $Q^2=S_{ep} x y>500\,\mathrm{GeV}^2$ with 
$0.01\le y \le 0.9$ and $\sqrt{S_{ep}}=300\,\mathrm{GeV}$ for HERA.
The GRV-94 NLO parton densities \cite{ref:grv94}, 
$m_c=1.5\,\mathrm{GeV}$, $\mu_F^2=Q^2+m_c^2$, and
$\varepsilon_c=0.06$ in the charm fragmentation function
(\ref{eq:peterson}) are used. Also shown are the individual NLO 
gluon- and quark-initiated contributions, the latter including the
virtual corrections, and the Born term obtained with NLO parton distributions.
\item[Fig.~3] 
The CC $\bar{D}$ meson production cross section
$e^-p\rightarrow \bar{D} X$ in NLO obtained from Eq.~(\ref{eq:unpolxsec})
as in Fig.~2 but now 
as a function of $x$ and integrated over two ranges of $z$: 
$0<z<1$ and $0.2<z<1$. Also shown is the Born contribution in each case, 
obtained with NLO parton distributions.
\item[Fig.~4] Comparison of the NLO $z$ integrated $\bar{D}$ 
meson production cross section $e^-p\rightarrow \bar{D} X$ 
as in Fig.~3 using the GRV-94 \cite{ref:grv94}
and MRST \cite{ref:mrst} distributions for two 
ranges of $z$: {\bf (a)} $0<z<1$ and {\bf (b)} $0.2<z<1$.  
\item[Fig.~5] As in Fig.~2 but now for longitudinally polarized
$e^-p$ collisions according to Eq.~(\ref{eq:polxsec}) and using the
GRSV `standard' set of polarized parton distributions \cite{ref:grsv}.
\item[Fig.~6] As in Fig.~3 but now for longitudinally polarized
$e^-p$ collisions according to Eq.~(\ref{eq:polxsec}) and using the
GRSV `standard' set of polarized parton distributions \cite{ref:grsv}.
\item[Fig.~7] The $z$ integrated polarized cross section
[{\bf (a)}: $0<z<1$,{\bf (c)}: $0.2<z<1$]  
as in Fig.~6 and the corresponding measurable spin asymmetry $A^D$ 
[{\bf (b)}: $0<z<1$,{\bf (d)}: $0.2<z<1$]
according 
to (\ref{eq:asym}). Also shown for comparison are the results 
for $e^+p\rightarrow D X$ and the ones obtained using the 
GRSV `valence' set of spin-dependent parton densities \cite{ref:grsv}.
The error bars denote the expected statistical accuracy $\delta A^D$
according to (\ref{eq:error}) for three different $x$ bins
assuming ${\cal{L}}=500\,\mathrm{pb}^{-1}$, $P_e=P_p=0.7$, and
$\varepsilon_{eff}^c=1$.
Two $x$ bins are chosen at equal logarithmic distance in the range 
$[x_{\min}\simeq 0.006,0.1]$ and one for $x>0.1$.
\end{description}
%

\newpage
\pagestyle{empty}
\hspace*{1.5cm}

\begin{center}
\hspace*{1.3cm}
\begin{picture}(250,90)(0,-50)
\ArrowLine(50,20)(100,40)
\Vertex(100,40){2.0}
\ArrowLine(100,40)(150,20)
\Photon(100,80)(100,40){3}{6}
\Text(50,10)[]{$p_s$, $m_s=0$}
\Text(150,10)[]{$p_c$, $m_c\neq 0$}
\Text(110,60)[]{$q$}
\Text(100,-20)[]{(a)}
\end{picture}
\end{center}
\begin{center}
\hspace*{0.722cm}
\begin{picture}(450,90)(0,-50)
\ArrowLine(60,0)(80,20)
\Line(80,20)(100,40)
\Vertex(100,40){2.0}
\ArrowLine(100,40)(170,40)
\Photon(80,80)(100,40){3}{6}
\Gluon(80,20)(125,20){-2}{6}
\Vertex(80,20){2.0}
\Text(60,-10)[]{$p_s$}
\Text(170,30)[]{$p_c$}
\Text(105,60)[]{$q$}
\Text(200,-20)[]{(b)}
\Text(125,10)[]{$k$}
\ArrowLine(260,0)(300,40)
\Vertex(300,40){2.0}
\Line(300,40)(335,40)
\ArrowLine(335,40)(370,40)
\Photon(280,80)(300,40){3}{6}
\Gluon(335,40)(350,10){-2}{6}
\Vertex(335,40){2.0}
\Text(260,-10)[]{$p_s$}
\Text(370,30)[]{$p_c$}
\Text(305,60)[]{$q$}
\Text(350,0)[]{$k$}
\end{picture}
\end{center}
\begin{center}
\hspace*{-1.cm}
\begin{picture}(550,90)(0,-50)
\ArrowLine(50,20)(78,31)
\Line(78,31)(100,40)
\Vertex(100,40){2.0}
\Line(100,40)(122,31)
\ArrowLine(122,31)(150,20)
\Photon(100,80)(100,40){3}{6}
\GlueArc(100,40)(25,-157,-23){-3}{6}
\Vertex(78,31){2.0}
\Vertex(122,31){2.0}
\Text(50,5)[]{$p_s$}
\Text(150,5)[]{$p_c$}
\Text(110,60)[]{$q$}
\Line(200,20)(211,25)
\ArrowLine(211,25)(238,35)
\Line(238,35)(250,40)
\Vertex(250,40){2.0}
\Photon(250,80)(250,40){3}{6}
\ArrowLine(250,40)(300,20)
\GlueArc(225,30)(15,-156,24){-2}{10}
\Vertex(211,25){2.0}
\Vertex(238,35){2.0}
\Text(200,5)[]{$p_s$}
\Text(300,5)[]{$p_c$}
\Text(260,60)[]{$q$}
\Text(250,-20)[]{(c)}
\ArrowLine(350,20)(400,40)
\Vertex(400,40){2.0}
\Photon(400,80)(400,40){3}{6}
\Line(400,40)(411,35)
\ArrowLine(411,35)(438,25)
\Line(438,25)(450,20)
\GlueArc(425,30)(15,-204,-24){-2}{10}
\Vertex(411,35){2.0}
\Vertex(438,25){2.0}
\Text(350,5)[]{$p_s$}
\Text(450,5)[]{$p_c$}
\Text(410,60)[]{$q$}
\end{picture}
\end{center}
\begin{center}
\begin{picture}(450,90)(0,-50)
\hspace*{0.722cm}
\Gluon(60,0)(80,20){-2}{6}
\ArrowLine(80,20)(100,40)
\Vertex(100,40){2.0}
\ArrowLine(100,40)(170,40)
\Photon(80,80)(100,40){3}{6}
\ArrowLine(125,20)(80,20)
\Vertex(80,20){2.0}
\Text(60,-10)[]{$p_g$}
\Text(170,30)[]{$p_c$}
\Text(105,60)[]{$q$}
\Text(200,-20)[]{(d)}
\Text(125,10)[]{$p_{\bar s}$}
\Gluon(260,0)(280,20){-2}{6}
\ArrowLine(300,40)(280,20)
\Vertex(300,40){2.0}
\ArrowLine(370,40)(300,40)
\Photon(280,80)(300,40){3}{6}
\ArrowLine(280,20)(325,20)
\Vertex(280,20){2.0}
\Text(260,-10)[]{$p_g$}
\Text(370,30)[]{$p_{\bar s}$}
\Text(305,60)[]{$q$}
\Text(325,10)[]{$p_c$}
\end{picture}
\end{center}
\begin{center}
{\bf{Fig.~1}}
\end{center}

\newpage
\renewcommand{\textfraction}{0}
\renewcommand{\topfraction}{1}

\vspace*{-3.5cm}
\hspace*{-2.7cm}
\epsfig{figure=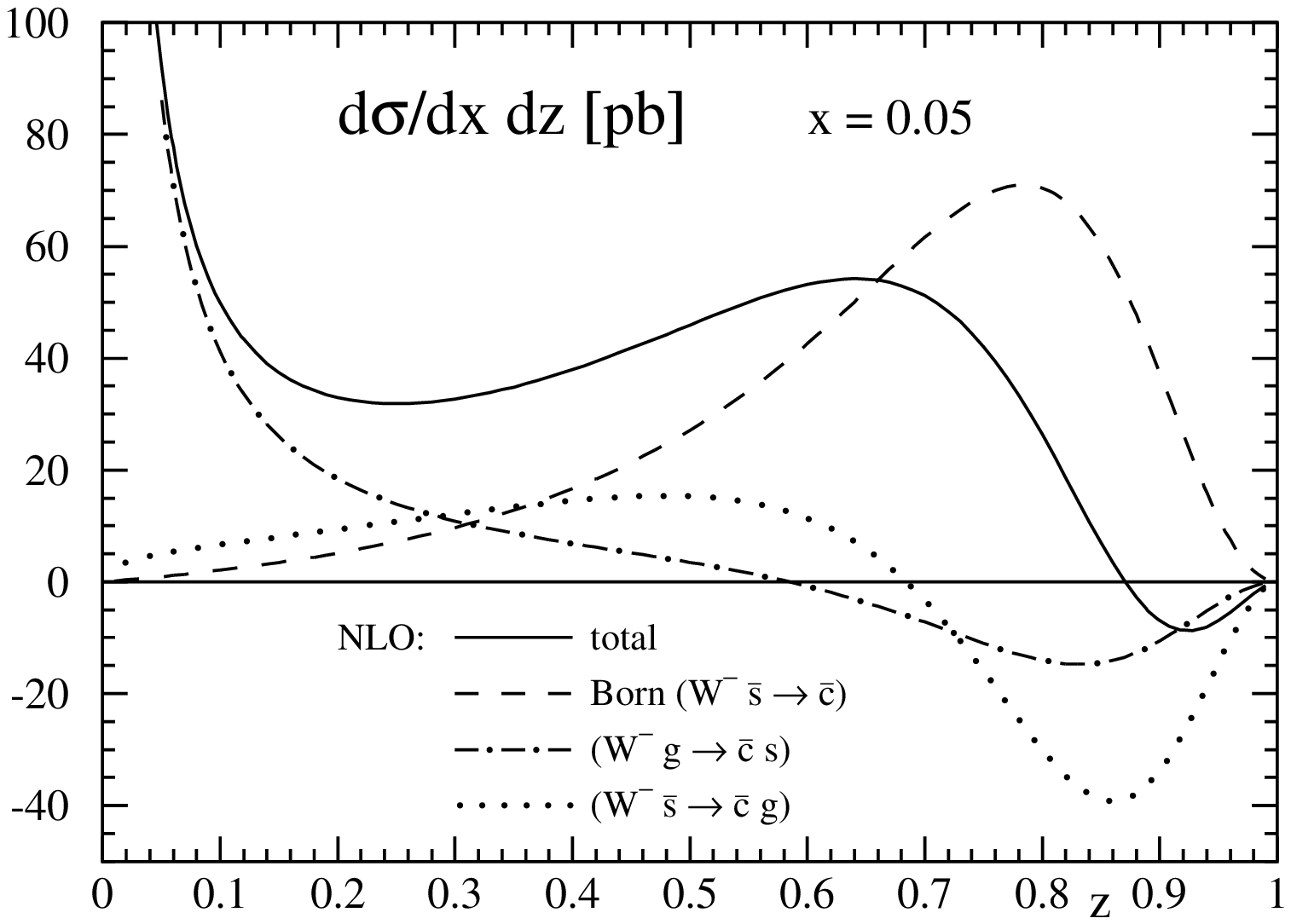,width=19cm}
\begin{center}
\vspace*{-3.7cm}
{\bf{Fig.~2}}
\end{center}
\vspace*{-1.9cm}
\hspace*{-2.cm}
\epsfig{figure=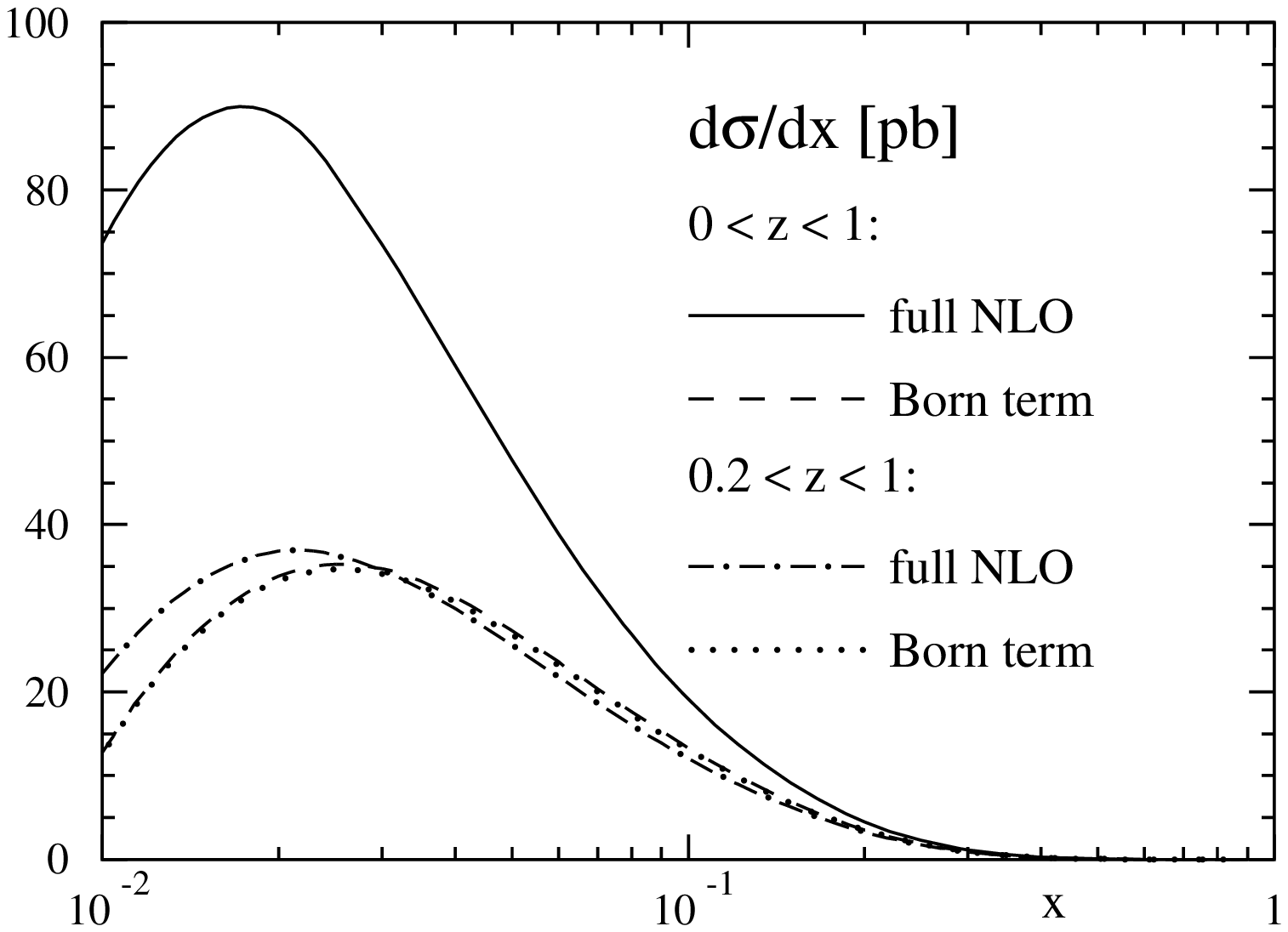,width=19cm}
\begin{center}
\vspace*{-3.7cm}
{\bf{Fig.~3}}
\vspace*{-2cm}
\end{center}

\newpage

\vspace*{-1.cm}
\hspace*{-1.cm}
\epsfig{figure=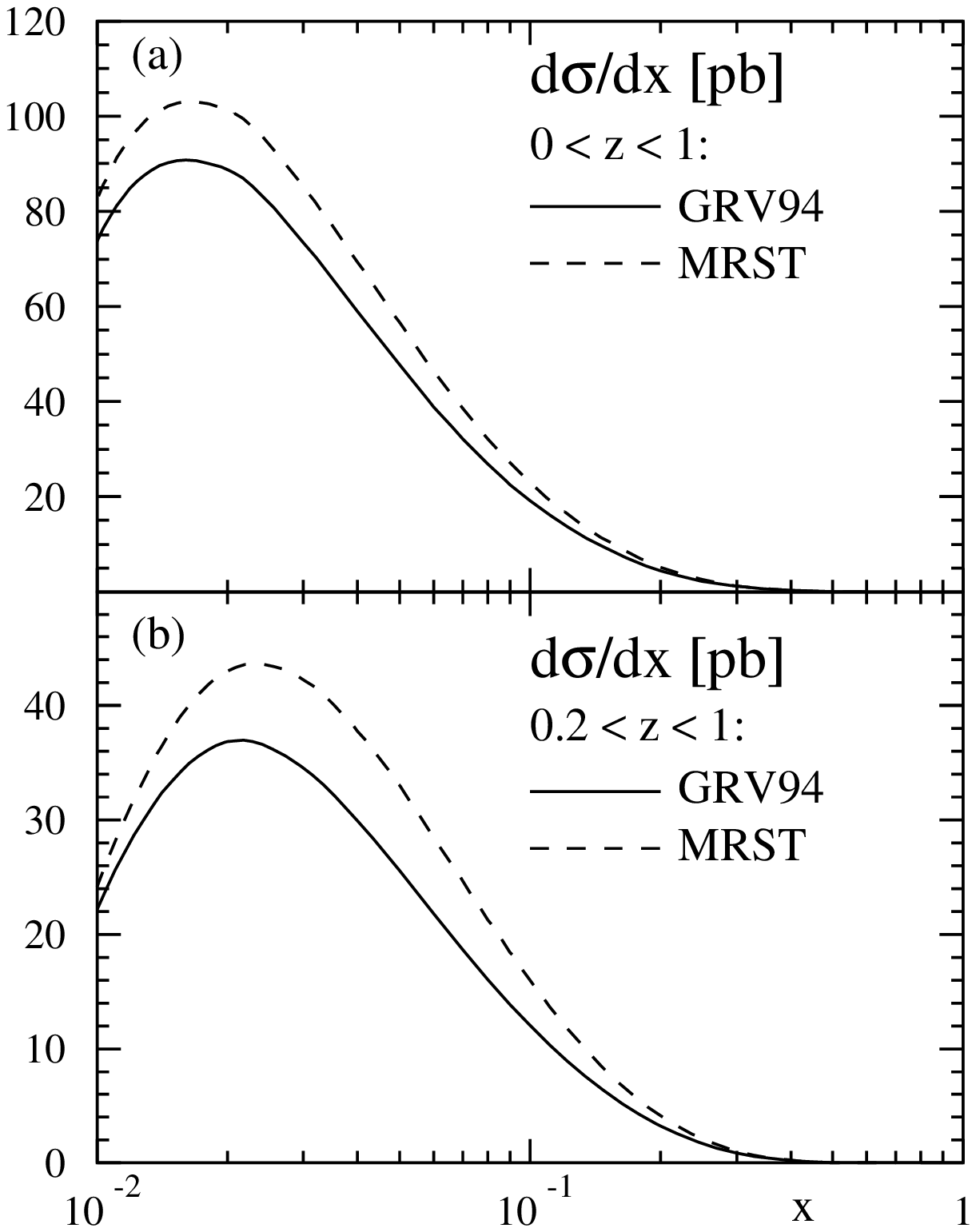}
\begin{center}
\vspace*{-2.3cm}
{\bf{Fig.~4}}
\vspace*{-2cm}
\end{center}

\newpage

\vspace*{-3.5cm}
\hspace*{-2.7cm}
\epsfig{figure=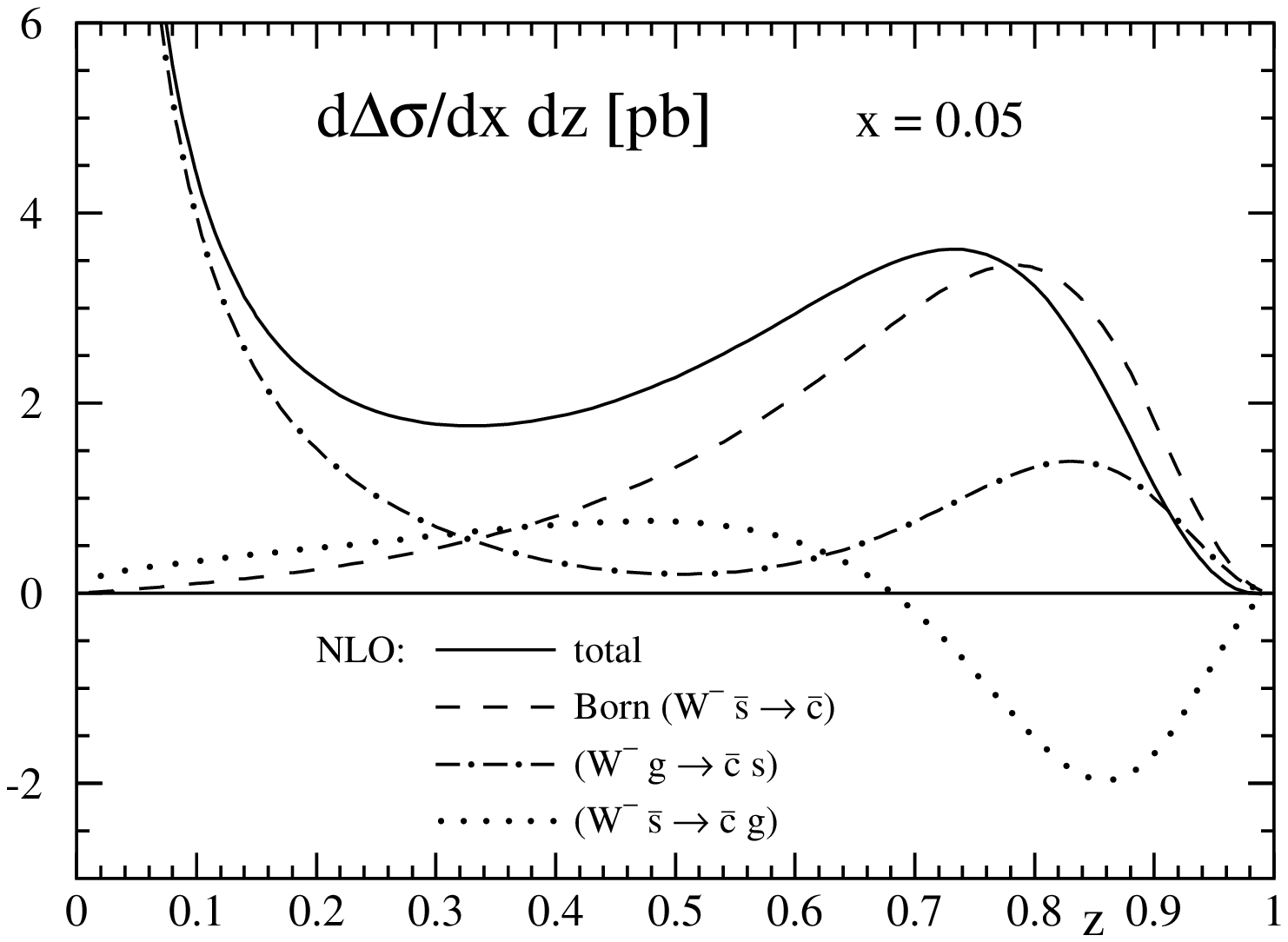,width=19cm}
\begin{center}
\vspace*{-3.7cm}
{\bf{Fig.~5}}
\end{center}
\vspace*{-1.9cm}
\hspace*{-2.cm}
\epsfig{figure=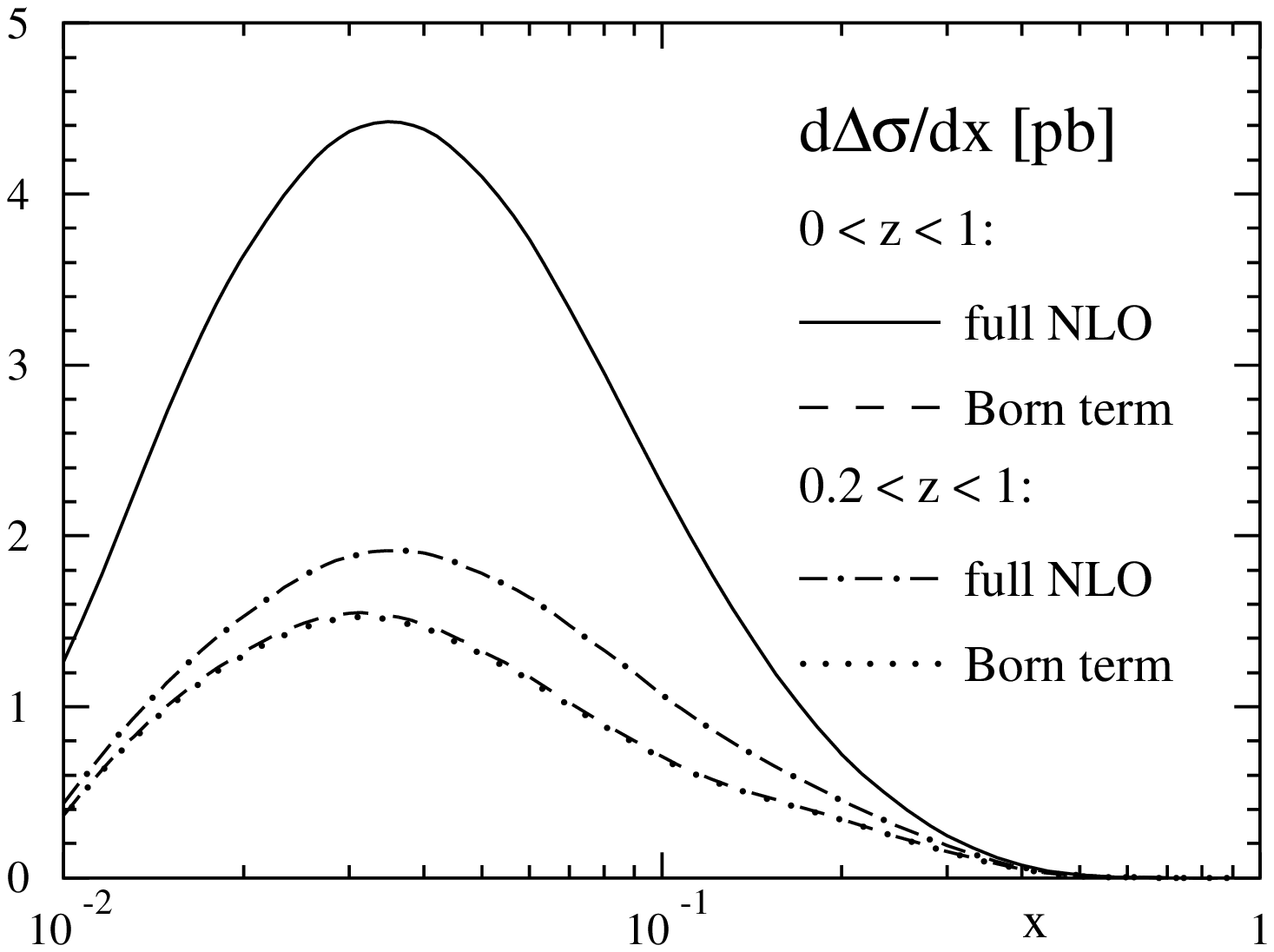,width=19cm}
\begin{center}
\vspace*{-3.7cm}
{\bf{Fig.~6}}
\vspace*{-2cm}
\end{center}

\newpage

\vspace*{-2.8cm}
\hspace*{-2.2cm}
\epsfig{figure=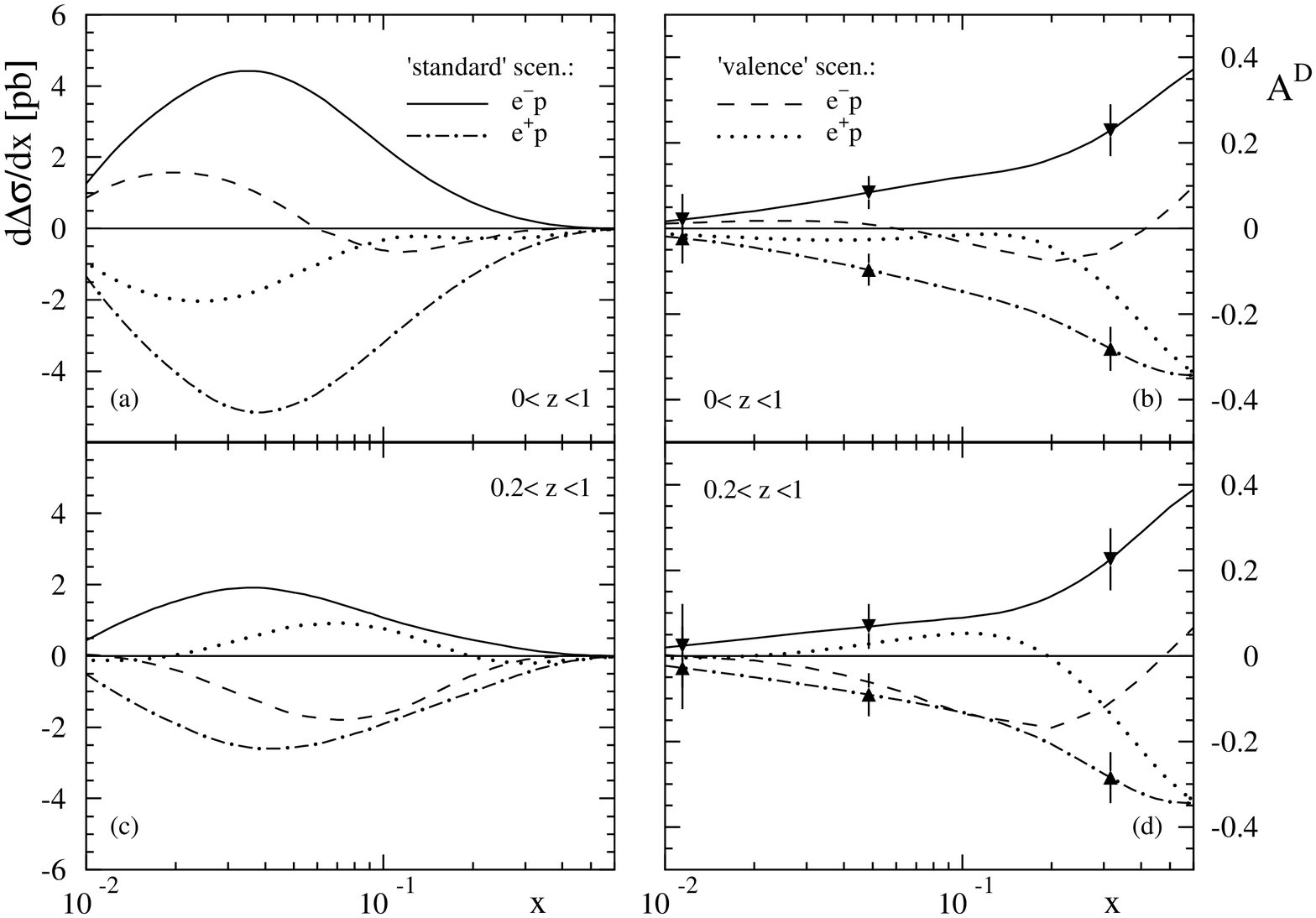,angle=90,width=18cm}
\begin{center}
\vspace*{-2.2cm}
{\bf{Fig.~7}}
\vspace*{-2cm}
\end{center}

\end{document}